%% file: preprint.tex
\documentclass[11pt]{article}

\usepackage{amssymb,amsmath,amsfonts,eurosym,geometry,ulem,graphicx,color,setspace,sectsty,comment,footmisc,caption,pdflscape,subfigure,array,placeins,adjustbox,booktabs,longtable,dirtytalk,tabularx,multirow,soul,threeparttable,float,rotating, svg, authblk, bbm, endnotes,tikz,standalone,listings,xcolor}
\usepackage{csquotes}
\usepackage{CJKutf8}
\usepackage{authblk}
\usepackage{enumitem}
\usetikzlibrary{arrows.meta,positioning,calc}

\usepackage{natbib}
 \bibpunct[, ]{(}{)}{,}{a}{}{,}%

\usepackage{bibunits}
\usepackage[hyperfootnotes=false, colorlinks=true, linkcolor=blue, urlcolor=blue, citecolor=blue]{hyperref}

\lstdefinestyle{promptstyle}{
  basicstyle   = \ttfamily\scriptsize,
  mathescape   = true,
  frame        = single,
  breaklines   = true,
  breakatwhitespace = true,
  postbreak    = \mbox{$\hookrightarrow$\space},
  escapeinside = {(*@}{@*)}
}

\newcommand{\zh}[1]{\begin{CJK*}{UTF8}{gbsn}#1\end{CJK*}}

\newcolumntype{L}[1]{>{\raggedright\let\newline\\arraybackslash\hspace{0pt}}m{#1}}
\newcolumntype{C}[1]{>{\centering\let\newline\\arraybackslash\hspace{0pt}}m{#1}}
\newcolumntype{R}[1]{>{\raggedleft\let\newline\\arraybackslash\hspace{0pt}}m{#1}}

\begin{document}

\begin{bibunit}[plainnat]

\title{Trusting AI in Competitive Markets}


\author{Jussi Keppo, Yuze Li, Gerry Tsoukalas, Nuo Yuan\thanks{Keppo (keppo@nus.edu.sg): National University of Singapore; Li (yuzeli@cuhk.edu.hk): Chinese University of Hong Kong; Tsoukalas (gerryt@bu.edu): Boston University; Yuan (nuo.yuan@cityu-dg.edu.cn): City University of Hong Kong (Dongguan).}}

\date{\today}

\maketitle

\begin{abstract}

People's trust in AI advice diverges as they use it, deepening for some and eroding for others. We study this divergence in oligopoly pricing, where advice cannot prove itself: rivals' responses decide whether it pays off. In a laboratory experiment, 273 sellers compete across 91 three-seller markets over 30 rounds; we vary the presence of AI pricing recommendations and the gender composition of the market (female-only, male-only, or mixed). We find that the gender composition of the market shapes how sellers learn from the advice, and where prices settle as a result. In female-only markets, recommendations raise prices by 29\% and profits by 39\%; in male-only and mixed-gender markets, they have no significant effect. A Non-Homogeneous Hidden Markov Model reveals a composition-specific dynamic association: profitable rounds predict rising adherence to the AI in female-only markets and declining adherence otherwise, a pattern consistent with learned trust and self-serving attribution. The pattern reverses what recent evidence on gender and AI would predict. We discuss implications for platform governance and regulatory oversight, which should focus not only on the algorithm but on the human side that shapes its effects.

\bigskip

\noindent\textbf{Keywords:} Algorithmic Pricing, Artificial Intelligence, Collusion, Gender, Generative AI, Human-AI Interaction.

\end{abstract}


\section{Introduction}\label{sec:intro}

AI now sits alongside workers in many jobs, recommending actions they are free to follow or ignore \citep{bick2026rapid, brynjolfsson2025generative}. What governs that choice is not fully understood \citep{logg2019algorithm}. The AI's track record explains part of it: reliance tends to track whether the advice has been right or wrong so far \citep{bastani2025improving, dietvorst2014algorithm}. But the record cannot be the whole story, because people who face the same advice end up relying on it to very different degrees \citep{liu2025find}. We ask how people learn to work with AI advice in competitive markets, where the track record depends on others' strategic responses: outcomes are clear, but what they prove about the advice is not. Understanding what drives this reliance matters because biases might emerge in AI-mediated decision-making contexts when people decide to defer to the machine \citep{huang2026beyond}.

Most of what we know about reliance on AI comes from people deciding alone, where advice is right or wrong on its own terms. Markets work differently: when a few sellers price against one another, the evidence on whether the advice works emerges from how they interact, not from the advice alone. The same recommendation can pay off in one market and backfire in another. Success is also hard to attribute: a seller who follows the advice and profits can credit the AI or their own judgment, and because pricing repeats, that reading feeds back into the market. A seller who credits the AI follows it more closely next round. That changes what rivals earn, and rivals draw their own conclusions from it. Small differences in how sellers interpret the same outcome can therefore compound over time, so two identical markets can end up in very different places, one at coordinated high prices, the other competed down.

Whether a market competes down or coordinates up may depend on who is in it. Gender shapes how people compete and coordinate in groups, and how they take to AI in the first place: women adopt generative AI less than men, a gap that holds across regions and persists even when access is equal \citep{otis2024global, humlum2025unequal}. Adoption is typically a one-time choice, made before the technology has proven anything. Reliance is different: it is revised every time the advice pays off or disappoints, and need not follow the same pattern. To see whether it does, we randomize the gender composition of markets and follow sellers' reliance round by round.

AI pricing recommendations are now standard on major marketplaces, reaching millions of sellers \citep{alizila2024}, and their tendency to sustain supra-competitive prices has drawn scrutiny from researchers and regulators \citep{calvano2020artificial, fish2024algorithmic, doj2025realpage}. This scrutiny has focused almost entirely on the algorithm. We argue that supra-competitive coordination is better understood as a property of the coupled human--AI system: the same algorithm can sustain coordination in one market and fail in another, depending on who is setting the prices and how they learn from the outcomes. We examine three questions: (i) whether AI pricing recommendations affect market outcomes, (ii) how that effect depends on the gender composition of the market, and (iii) what behavioral dynamic underlies the difference.

To answer these questions, we conduct a laboratory experiment at a university in Singapore, implementing a repeated Bertrand pricing game with symmetrically differentiated products, in which participants act as sellers competing in three-seller markets. Each market (or ``cluster'') consists of three sellers, each aiming to maximize long-run cumulative profit over multiple rounds. In each round, sellers observe their own previous-round price, market share, profit, and cumulative profit, as well as aggregate market statistics (minimum, average, and maximum prices) from their cluster. The experiment uses a $2 \times 3$ between-subjects factorial design that varies by the presence of AI pricing recommendations and cluster gender composition. First, each session was randomly assigned to either (i) no AI pricing recommendations (control), in which sellers did not receive AI-generated price recommendations, or (ii) AI pricing recommendations (treatment), in which sellers received an AI-generated price recommendation at the beginning of each round that they could follow or ignore, so that every participant in a session faced the same AI condition. Second, within each session, participants were randomly grouped by gender into clusters of one of three compositions: (i) female-only, (ii) male-only, or (iii) mixed-gender. The cluster's gender composition and sellers' individual identities were not revealed to participants, and they did not communicate with one another.

Before conducting the main experiment, we first characterize the AI's pricing behavior in a benchmark simulation where three GPT-5 mini agents play the Bertrand pricing game against each other. The AI consistently recommends prices averaging 52\% above the static Nash equilibrium, tightly clustered across agents (within-round range of only \$0.18). This supra-competitive pricing coordination emerges endogenously from the LLM's own reasoning \citep{fish2024algorithmic, keppo2026fragility}. The AI thus functions as a coordination-facilitating anchor that provides similar supra-competitive price signals to all sellers in the market.

We find that, pooled across all market compositions, the presence of AI pricing recommendations significantly raises the average market price ($p < 0.001$); the pooled profit effect is positive but does not reach significance ($p = 0.124$). In the absence of AI pricing recommendations, average market prices settle at approximately 19\% above the static Nash equilibrium. With AI pricing recommendations, market prices rise to approximately 30\% above Nash.

This aggregate effect masks substantial heterogeneity across market compositions. The significant pooled price effect is driven entirely by female-only markets, where AI pricing recommendations raise the average market price by approximately 29\% ($p < 0.001$). Aggregate cluster profit in these markets rises by approximately 39\% ($p < 0.001$), with effects sustained across rounds. In male-only and mixed-gender markets, AI pricing recommendations have no significant effect on either prices or profits. This pattern presents a puzzle: although the AI initially recommends supra-competitive prices across all compositions, market-level responses diverge sharply.

To characterize the behavioral dynamic, we estimate a Non-Homogeneous Hidden Markov Model (NHMM) of individual AI adherence separately for each gender composition. The model classifies each seller-round into a high-adherence state (priced close to the AI recommendation) or a low-adherence state (priced away from it), with transitions between states shaped by prior market feedback. We find that higher profits predict stronger AI adherence in female-only markets and weaker adherence in male-only markets. In female-only markets, profitable adherence is associated with a positive feedback pattern in which prices converge toward the AI's supra-competitive recommendation. In male-only and mixed-gender markets, the same success is associated with subsequent deviation, consistent with coordination unravelling and the AI's effect being neutralized.

Our work contributes to several literatures. Its central contribution is to the algorithmic-pricing literature, which has studied supra-competitive coordination mainly as a property of algorithms: autonomous agents can sustain it without explicit communication \citep{calvano2020artificial, fish2024algorithmic, keppo2026fragility}. When humans remain in the pricing loop, however, findings diverge: AI recommendations raise prices in some markets and lower them in others. Existing accounts have linked this divergence to algorithm features such as punishment severity \citep{hunold2024algorithmic}, the form of delegation \citep{normann2025delegate}, and recommendation bias \citep{rebholz2025advice}, and to market features such as size and the share of algorithmic sellers \citep{werner2026algorithmic}. We offer a different explanation: coordination is a property of the coupled human--AI system, not the algorithm alone. The same adaptive algorithm enters different feedback paths depending on the composition of the market, producing different recommendation sequences and different equilibrium outcomes. Algorithm-side oversight, such as restricting the data an algorithm may access \citep{doj2025realpage}, is therefore insufficient; the margin that warrants attention is the adherence the algorithm elicits. We show that the gender composition of human sellers is one dimension that shapes this adherence, a dimension absent from existing algorithm- and market-based accounts.

We build on the literature on human--AI reliance dynamics, which shows how each human's reliance on AI evolves with experience and context. When people see algorithmic advice work, compliance with it rises over rounds \citep{bastani2025improving}. Yet reliance can also erode once people watch an algorithm err \citep{dietvorst2014algorithm}. Even without visible errors, reliance is hard to calibrate, because people misjudge their own competence and defer at the wrong times \citep{fugener2022cognitive}, and the decision context can block the feedback needed to learn whether the machine is better \citep{devericourt2025your}. Delegation offers an organizing lens for this relationship \citep{baird2021next}, and recent work models its trajectory as a latent process, finding that reliance tends to polarize across decision-makers as some deepen it while others pull back \citep{liu2025find}. What this literature has not yet modeled is how reliance evolves in markets, where sellers' choices interact strategically and small differences in interpretation compound into divergent collective outcomes. We provide that account, tracing how reliance evolves and how it co-varies with the market outcomes that follow. We argue that the direction it takes depends on the composition of the market.

Research on gender and collective behavior shows that gender shapes how a group competes and coordinates, with all-female groups tending to collaborate more and price less aggressively \citep{ma2021women, apesteguia2012impact}. These tendencies reflect well-documented differences in competitive behavior: women shy away from competition while men are overconfident \citep{niederle2007women}, a gap that widens among same-gender peers: men signal toughness, women signal cooperativeness \citep{charness2011gender}. Closest to our setting, advice interacts with gender inside competition itself: when an experienced advisor counsels entry into a tournament, women are more likely to follow advice to hold back, and men more likely to follow advice to push forward \citep{brandts2015impact}. This evidence comes from settings without AI. We extend this to strategic markets where an AI recommends prices, asking how gender composition shapes the competition among sellers and the market outcomes that follow.

A growing literature documents a gender gap in AI adoption: women take up generative AI less than men, even when access is equal \citep{otis2024global, humlum2025unequal}. Women are more skeptical of AI and more concerned about its risks \citep{borwein2026explaining}, echoing a longer pattern in which gender shapes how people take up new technology \citep{venkatesh2000men}. Yet these studies measure a single adoption decision, not the repeated reliance that builds as people use AI again and again in strategic groups. Bringing together gender's effect on group competition and its effect on engagement with AI, we find that female-only markets coordinate on the AI's supra-competitive price recommendations, while male-only and mixed-gender markets do not. Despite this lower baseline adoption, sellers in female-only markets adhere more closely to those recommendations as profits accumulate, while those in male-only markets increasingly deviate. The two literatures closest to our question make opposite predictions. The adoption and skepticism evidence predicts that female sellers should discount the AI's recommendations. Classic evidence on attribution and overconfidence predicts the opposite, since men convert success into confidence in their own judgment while women credit it elsewhere \citep{beyer1990gender, barberodean2001boys, shastry2020luck, coffman2024stereotypes}. Our experiment tests these predictions against each other: market feedback favors the attribution account, and replayed round after round, this asymmetry is associated with whether an entire market ends up at supra-competitive prices.

The remainder of the paper proceeds as follows. Section~\ref{sec:design} describes the experimental design. Section~\ref{sec:data_empirical} presents the data and empirical strategy. Section~\ref{sec:results} reports the empirical results. Section~\ref{sec:discussion} discusses the behavioral mechanism, policy implications, and limitations for future research.

\section{Experimental Design}\label{sec:design}

We conduct a laboratory experiment in repeated Bertrand markets with differentiated products.

\subsection{Market Environment}\label{sec:market}

Each market consists of $n = 3$ sellers playing a repeated pricing game with symmetrically differentiated products. In each round, each seller simultaneously chooses a price for its product. Sellers do not know in advance how many rounds the game will last.

Demand for each seller follows a multinomial logit specification, in which consumers have idiosyncratic preferences across sellers so that a lower-priced seller captures more demand but not all of it \citep{anderson1992discrete, caplin1991aggregation}. Given the price vector $\mathbf{p} = (p_1, p_2, p_3)$, seller $i$'s market share is:
\begin{equation}\label{eq:demand}
    s_i(\mathbf{p}) = \frac{\exp\!\bigl((a - p_i)/\mu\bigr)}{\exp(a_0/\mu) + \sum_{j=1}^{n} \exp\!\bigl((a - p_j)/\mu\bigr)},
\end{equation}
where $a = 2.0$ is the product value index (common to all sellers), $a_0 = 0.0$ is the outside-option value, and $\mu = 0.25$ governs the degree of differentiation.

Each seller has constant marginal cost $c = 1.0$ and earns per-round profit $\pi_i(\mathbf{p}) = (p_i - c) \cdot s_i(\mathbf{p})$. Sellers aim to maximize cumulative profit across rounds.

Two theoretical benchmarks anchor our analysis. The Bertrand-Nash equilibrium price is $p^* = 1.37$, the symmetric price at which no seller can unilaterally improve profit. The joint-monopoly price is $p^m = 2.00$, the price that maximizes joint industry profit. (A non-binding price ceiling of $\gamma \cdot p^m$ with $\gamma \sim \text{Uniform}(1.5, 2.5)$, yielding \$3.00 to \$5.00, was enforced each session to rule out extreme prices \citep{fish2024algorithmic}; observed prices were well below this ceiling.) Prices above $p^*$ and approaching $p^m$ indicate supra-competitive pricing.

\subsection{Treatment Conditions}\label{sec:treatments}

Our experiment uses a between-subjects $2 \times 3$ factorial design, crossing AI pricing recommendations with cluster gender composition. We randomize AI treatment across experimental sessions and gender composition across clusters within each session. Each participant experiences exactly one condition, ruling out within-participant spillovers and yielding a cluster-level panel of repeated decisions.
\paragraph{Control (no AI).} In the control condition, participants make all pricing decisions independently. At the start of each round, they observe the previous round's own price, market share, profit, and cumulative profit, together with their cluster's aggregate market statistics (the minimum, average, and maximum prices). They are also shown the product description, their marginal cost, and the market structure. They do not know the identity of the other two sellers. Because each market has exactly three sellers, the reported minimum, average, and maximum prices allow a seller to recover the other two sellers' prices as numerical values, though not which competitor set which price. They receive no AI-generated recommendations.

\paragraph{Treatment (AI pricing recommendations).} In the treatment condition, in addition to all the information shown in the control condition, participants receive an AI-generated price recommendation at the start of each round. We disclose the source to participants: the instructions state that the recommendation is produced by GPT-5 mini, a large language model. The AI pricing recommendation is a single numerical price; participants are free to follow or ignore it. The recommendation is presented as a bare number with no accompanying rationale or explanation, and participants cannot interact with the AI beyond observing its recommended price: they cannot query it, request justification, or modify the prompt. This design is deliberately minimal, isolating the effect of the recommendation itself from any effect of conversational interaction with an AI system.

Although the treatment interface presents only a single number, two features distinguish this from a generic numerical nudge. First, the recommendation is adaptive: the LLM observes each market's pricing history and updates its recommendation round by round, so the signal co-evolves with the market. A static nudge, by contrast, cannot respond to what participants actually do. Second, the source is disclosed as AI: participants know the number comes from a large language model, not from a formula, an expert, or a random process. Source labeling activates prior beliefs about AI competence and reliability \citep{logg2019algorithm, dietvorst2014algorithm}, and the attribution and trust dynamics we document in Section~\ref{sec:mechanism} run through those beliefs. Replacing the AI label with a non-AI source would hold the numerical recommendation constant but alter the psychological channel through which sellers interpret success and failure, precisely the channel our results show matters.

To produce these recommendations, the LLM receives the same information set as the human participant, comprising the previous round's own price, market share, profit, and cumulative profit, the aggregate market statistics (minimum, average, and maximum prices), and the product's marginal cost. Like the human, it is not shown competitors' individual prices or profits. Our prompt instructs the LLM to ``set prices that maximize your own profit in the long run,'' provides the market history, and requests a price along with a rationale and forward-looking strategy (the full prompt appears in Appendix~\ref{sec:prompt_appendix}). Two design choices are deliberate: (i) the LLM receives no information about the Nash or monopoly benchmarks or the underlying demand parameters ($a$, $a_0$, $\mu$), and is free to recommend any price within the enforced bounds; and (ii) the prompt contains no language encouraging or discouraging coordination with rivals. These choices keep our setup generic: we study what an off-the-shelf LLM produces when asked to maximize long-run profit in a competitive market. Table~\ref{tab:info_set} summarizes the information available to each group.

\begin{table}[!htbp]
\centering
\caption{Information Set by Treatment Condition}
\label{tab:info_set}
\begin{threeparttable}
\begin{tabular}{lcc}
\toprule
\textbf{Information Type} & \textbf{Control} & \textbf{Treatment} \\
\midrule
Previous round's own price, market share, profit, and cumulative profit & \checkmark & \checkmark \\
Previous round's market statistics (min/avg/max price) & \checkmark & \checkmark \\
Product description, marginal cost, and market structure & \checkmark & \checkmark \\
AI recommended price & & \checkmark \\
\bottomrule
\end{tabular}
\end{threeparttable}
\end{table}

\paragraph{Cluster gender composition.}\label{sec:gender_comp} We construct clusters with controlled gender compositions, following the randomization strategy of \citet{hoogendoorn2013impact}. Before each session, the experimenter specifies the number of female-only and male-only clusters to form. After all participants complete onboarding (which includes a gender question), the clustering algorithm proceeds in three steps: (i) it randomly draws enough female participants to fill the pre-specified number of female-only clusters; (ii) it randomly draws enough male participants to fill the pre-specified number of male-only clusters; (iii) all remaining participants, both female and male, are pooled and randomly formed into mixed-gender clusters of three, balancing genders within each cluster (two women and one man, or one woman and two men, depending on the pool). Conditional on gender composition, individual-to-cluster assignment is random, and AI treatment is randomly determined at the session level.

Crucially, we never disclose the gender composition to participants; they know only that they compete against two other sellers. This avoids demand effects in which knowledge of a counterpart's gender directly shifts pricing behavior, so our design isolates the role of composition through actual interaction rather than belief priming. The procedure yields three composition types: (i) female-only, (ii) male-only, and (iii) mixed-gender. Within each, we estimate the AI treatment effect separately, then contrast across them.

\subsection{Procedures}\label{sec:implementation}

We conducted the experiment at a public research university in Singapore, with approval from the university's Institutional Review Board (IRB). We recruited 273 participants from the university's student population, drawing from both master's and undergraduate students. Every participant who completed the experiment received a 10-SGD gift voucher, and the top earner in each session received an additional 10-SGD voucher, providing a direct financial incentive to compete on profit. The top-earner bonus introduces a rank-order tournament component. Because the bonus is identical across treatment and control within every gender composition, it cannot generate the treatment $\times$ composition interaction that is our key result; however, it may affect pricing levels and could interact with gender-specific responses to competition \citep{niederle2007women}. We return to this when interpreting the composition-specific mechanisms in Section~\ref{sec:mechanism}.

The experiment was run across seven sessions, each randomly assigned to a single AI-treatment level: four ran the AI treatment and three ran the control. Sessions were held in a large university classroom, with substantial spacing between participants to prevent communication and visual contact. Each participant interacted with a custom-built web platform we developed for this experiment. A typical session lasted approximately three hours.

To ground the abstract pricing game in a practical setting, we framed the experiment as an e-commerce marketplace where retailers compete on price to attract consumer demand, a context familiar to most participants from everyday online shopping. Each participant played the role of a retailer on Taobao, a major online marketplace, selling Mountain Peak Coffee Beans (Single Origin Ethiopian, 12oz bag) against two other retailers selling the same product. Participants were told that all three retailers sell an identical product and compete purely on price; the differentiation in our demand model comes only from consumers' idiosyncratic preferences over sellers (the logit term in Section~\ref{sec:market}), which participants cannot observe, so the framing is consistent with the demand structure we analyze. All three retailers in a market faced the same wholesale cost of \$1 per bag, corresponding to the marginal cost $c$ in Section~\ref{sec:market}. (In keeping with the Taobao framing, the interface displayed in-game prices in yuan; throughout the paper we report in-game amounts with the \$ symbol, denoting the same experimental currency unit. Vouchers were paid in SGD.) In each round, participants set a retail price for their product. Participants were instructed that the objective was to maximize cumulative profit over the course of this repeated pricing game. Our instructions conveyed only the broad economic intuition that lower prices would attract more consumer demand; we did not provide the underlying demand parameters, the theoretical benchmarks, or any language suggesting coordination or collusion among sellers. Appendix Figure~\ref{fig:ui} shows the interface for the control and treatment conditions.

Each session proceeded through five stages. (i) \textbf{Consent and instructions:} after arriving, participants reviewed the consent form and on-screen instructions describing the e-commerce scenario; we then walked through a screenshot demonstration of the interface for their assigned condition (Appendix Figure~\ref{fig:ui}), explained the payment scheme, and confirmed comprehension before the rounds began. (ii) \textbf{Onboarding survey:} each participant completed a pre-experiment survey collecting demographics (including gender). (iii) \textbf{Cluster assignment:} our clustering algorithm randomly assigned each participant to a cluster of the designated gender composition, following the procedure described in Section~\ref{sec:treatments}; cluster and seller identifiers remained hidden from participants. (iv) \textbf{Pricing rounds:} participants played 30 rounds of the pricing game. Each round was simultaneous-move: all three sellers in a cluster submitted their prices along with a free-text rationale for their decision (Appendix Figure~\ref{fig:rationale_ui}), and the round's results were revealed before the next round began. There was no per-round time limit, and not disclosing the number of rounds precluded backward induction by participants. (v) \textbf{Debrief and payment:} after the final round, participants were paid as described above and dismissed.


\section{Data and Empirical Strategy}\label{sec:data_empirical}

\subsection{Data}\label{sec:data}

Our 273 participants are evenly split by gender (50.9\% female, 48.4\% male, 0.7\% prefer not to say), predominantly Asian (90.5\%), with a mean age of 21.5 years (SD = 4.1). The majority (67.8\%) major in business administration or economics, with the remainder drawn from engineering, computer science, psychology, and other fields. Among those who responded, 37.0\% report prior experience selling on e-commerce marketplaces, bringing first-hand exposure to the pricing decisions our experiment studies. Participants also rated their prior trust in GPT on a 1--5 scale (mean = 3.06, SD = 0.80), with no significant difference across gender compositions. Table~\ref{tab:demographics} reports the full demographic breakdown.

\begin{table}[H]
\centering
\caption{Participant Demographics}
\label{tab:demographics}
\begin{threeparttable}
\renewcommand{\arraystretch}{0.9}
\begin{tabular}{llc}
\toprule
\textbf{Characteristic} & \textbf{Category} & \textbf{Overall \%} \\
\midrule
\multicolumn{3}{l}{Gender} \\[2pt]
& Female & 50.9 \\
& Male & 48.4 \\
& Prefer not to say & 0.7 \\[2pt]
\multicolumn{3}{l}{Ethnicity} \\[2pt]
& Asian & 90.5 \\
& White & 4.4 \\
& Other/Prefer not to say & 5.1 \\[2pt]
\multicolumn{3}{l}{Undergraduate Major} \\[2pt]
& Business Administration/Business & 54.6 \\
& Economics & 13.2 \\
& Engineering & 11.0 \\
& Computer Science & 6.2 \\
& Psychology & 3.3 \\
& Other & 11.7 \\[2pt]
\multicolumn{3}{l}{E-Commerce Selling Experience} \\[2pt]
& Yes & 37.0 \\
& No & 63.0 \\[2pt]
\multicolumn{3}{l}{Prior Trust in GPT (1--5 scale)} \\[2pt]
& 1 (very low) & 2.3 \\
& 2 & 19.0 \\
& 3 & 51.9 \\
& 4 & 24.1 \\
& 5 (very high) & 2.8 \\
& \multicolumn{2}{l}{Mean = 3.06, SD = 0.80} \\
\bottomrule
\end{tabular}
\end{threeparttable}
\end{table}

The 273 participants formed 91 three-seller clusters, each playing 30 rounds. After restricting to cluster-rounds with valid price and profit data for all three sellers, our analysis sample comprises 2{,}430 cluster-round observations. The 91 clusters split into 30 female-only, 30 male-only, and 31 mixed-gender, with 42 in the control condition and 49 in the AI-treatment condition. Appendix Table~\ref{tab:cluster_dist} reports the full cross-tabulation by treatment and gender composition. Female-only and male-only clusters are perfectly balanced across conditions (15 treatment, 15 control each); mixed-gender clusters are less balanced (19 treatment, 12 control) because three of the four sessions that contain mixed-gender clusters were assigned to treatment. Appendix~\ref{sec:balance} verifies covariate balance: most cluster-level covariates are comparable across treatment and control within each composition, and the few significant differences are absorbed by the controls in all regression specifications.

Because AI treatment is randomized at the session level, the inferential foundation rests on the seven sessions as independent units. Appendix Table~\ref{tab:session_layout} presents the full session-by-treatment-by-composition layout. Not all compositions appear in every session: female-only clusters are present in all seven sessions, male-only clusters in five (Sessions 3--7), and mixed-gender clusters in four (Sessions 1--4). This structure determines the effective number of session-level clusters available for session-robust inference in each composition subsample (Appendix~\ref{sec:session_clustering}). Treatment does not vary within session, so it is collinear with session indicators and the treatment specifications cannot include session fixed effects. The identifying assumption is therefore that session assignment to treatment is as-good-as-random. This leaves a finite-session risk that an unobserved baseline difference across sessions contributes to the estimated effect. We address this risk by reporting the full session layout and using session-level inference.

Table~\ref{tab:variables} describes the key variables. Our two cluster-round outcomes are the cluster average price (the mean across the three sellers' prices) and the cluster aggregate profit (the sum of their per-round profits). Three cluster-level indicators capture the experimental conditions. The Treatment indicator equals 1 for clusters that received AI-generated price recommendations and 0 for control clusters that did not. FemaleOnly equals 1 for all-female clusters; Mixed equals 1 for mixed-gender clusters; male-only clusters are the omitted baseline.

\begin{table}[!htbp]
\centering
\caption{Variable Definitions}
\label{tab:variables}
\begin{threeparttable}
\begin{tabular}{lll}
\toprule
\textbf{Variable} & \textbf{Level} & \textbf{Definition} \\
\midrule
\multicolumn{3}{l}{Dependent Variables} \\[3pt]
Average Price & Cluster--round & Mean price across the 3 sellers in cluster $i$, round $t$ \\
Aggregate Profit & Cluster--round & Sum of profits across the 3 sellers in cluster $i$, round $t$ \\[6pt]
\multicolumn{3}{l}{Treatment and Composition} \\[3pt]
Treatment & Cluster & $= 1$ if the cluster receives AI price recommendations \\
FemaleOnly & Cluster & $= 1$ if all 3 sellers are female \\
Mixed & Cluster & $= 1$ if the cluster contains both genders \\
Female Proportion & Cluster & Share of female sellers in the cluster (mixed only) \\[6pt]
\multicolumn{3}{l}{Controls} \\[3pt]
Average Age & Cluster & Mean age of the 3 sellers \\
Age Diversity & Cluster & Sample standard deviation of sellers' ages \\
Ethnicity/Race Diversity & Cluster & Blau index of sellers' ethnicity/race categories \\
Major Diversity & Cluster & Blau index of sellers' undergraduate majors \\
E-Commerce Prevalence & Cluster & Proportion of sellers with prior e-commerce selling experience \\
\bottomrule
\end{tabular}
\begin{tablenotes}
\footnotesize
\item Notes. Male-only is the omitted reference category for gender composition. The Blau index is defined as $1 - \sum_k p_k^2$, where $p_k$ is the proportion of sellers in category $k$; it equals 0 when all sellers share the same category and approaches 1 with maximum diversity.
\end{tablenotes}
\end{threeparttable}
\end{table}

\subsection{Empirical Strategy}\label{sec:econometric}

We test our research questions in three steps. First, we estimate the average effect of AI pricing recommendations on market prices and profits, pooled across all gender compositions (RQ1), using the following panel regression at the cluster--round level:
\begin{equation}\label{eq:pooled}
    Y_{it} = \alpha + \beta_1 \, \text{Treatment}_i + \beta_2 \, \text{FemaleOnly}_i + \beta_3 \, \text{Mixed}_i + \boldsymbol{\gamma}' \mathbf{X}_i + \delta_t + \varepsilon_{it},
\end{equation}
where $Y_{it}$ is the outcome (cluster average price or aggregate profit) for cluster $i$ in round $t$; $\text{Treatment}_i$ equals 1 for clusters receiving AI pricing recommendations and 0 for control clusters; $\text{FemaleOnly}_i$ and $\text{Mixed}_i$ are gender-composition indicators with male-only as the omitted baseline; $\mathbf{X}_i$ is the vector of cluster-level controls listed in Table~\ref{tab:variables}; and $\delta_t$ are round fixed effects. The coefficient $\beta_1$ identifies the average treatment effect of AI pricing recommendations on the outcome. We cluster standard errors at the cluster level to account for within-cluster dependence across rounds.

Second, to test whether and how gender composition moderates AI's effect on market outcomes (RQ2), we estimate the same specification separately within each gender-composition subsample $g \in \{\text{female-only, male-only, mixed-gender}\}$:
\begin{equation}\label{eq:subsample}
    Y_{it}^{(g)} = \alpha^{(g)} + \beta_1^{(g)} \, \text{Treatment}_i + \boldsymbol{\gamma}^{(g)\prime} \mathbf{X}_i + \delta_t^{(g)} + \varepsilon_{it}^{(g)},
\end{equation}
where superscript $(g)$ indexes the composition subsample, and the remaining variables are as defined in \eqref{eq:pooled}. For the mixed-gender subsample, we additionally include the female proportion within the cluster as a covariate to absorb within-subsample variation in gender share. Comparing $\beta_1^{(g)}$ across the three subsamples reveals whether AI's effect on market outcomes varies by gender composition.

Third, to characterize the behavioral dynamic through which gender composition may produce these heterogeneous AI effects (RQ3), we use a Non-Homogeneous Hidden Markov Model (NHMM) to study individual AI adherence dynamics, following the framework of \citet{liu2025find}. We measure participant $i$'s adherence to the AI's price recommendation in round $t$ by the absolute deviation $d_{it} = |p_{it} - p_{it}^{\mathit{AI}}|$, where $p_{it}$ denotes the participant's chosen price and $p_{it}^{\mathit{AI}}$ the AI's recommendation. The model has three components: (i) two latent states, a high-adherence state ($s = 0$, in which a participant prices near the AI's recommendation) and a low-adherence state ($s = 1$, in which they deviate substantially) --- two states correspond to the theoretically motivated binary distinction between following and ignoring the AI's recommendation, as finer partitions (e.g., partial adherence) lack a clear behavioral interpretation in our setting where the AI issues a single point recommendation each round and the seller either prices near it or does not --- (ii) state-specific emission distributions that model how $d_{it}$ is distributed in each state, capturing the typical magnitude and variability of price deviations from the AI; and (iii) logistic transitions that allow recent market outcomes to shift state probabilities in the next round.

In each round $t$, each participant resides in one of the two latent states. By the end of round $t$, they observe the round's outcome (their own price, profit, market aggregates, and the AI's recommendation), which may drive them to transition to a different state in round $t+1$. We model the emission as Gaussian:
\begin{equation}\label{eq:nhmm_emission}
d_{it} \mid S_{it} = s \sim N(\mu_s, \sigma_s^2),
\end{equation}
where the state-specific mean $\mu_s$ captures the typical magnitude of deviation in state $s$ and $\sigma_s$ captures its variability, with $\mu_0 < \mu_1$ by ordering convention. Because $d_{it} \geq 0$, the Gaussian places some probability mass on negative values; however, the large separation between states makes the decoded state paths robust to this boundary mismatch, and the Gaussian emission is standard in HMM studies of bounded observables \citep[e.g.,][]{liu2025find}.

Transitions between states are driven by recent market outcomes. Accordingly, each participant $i$'s transition propensity out of state $j$ at round $t$, $\eta_{i,j,t}$, is:
\begin{equation}\label{eq:nhmm_propensity}
\eta_{i,j,t} = \boldsymbol{\beta}_j' \mathbf{x}_{i,t} + \xi_i,
\end{equation}
where $\mathbf{x}_{i,t}$ collects lagged own outcomes (price, market share, profit, cumulative profit), lagged aggregate market prices, and the AI recommended price. Since participants in different states may respond differently to the same market signals, we define $\boldsymbol{\beta}_j$ as a state-dependent coefficient vector representing how each covariate shifts the transition propensity out of state $j$. The participant-specific factor $\xi_i \sim N(0, \sigma_\xi^2)$ is a random-effect term that absorbs unobserved heterogeneity in participants' baseline propensity toward AI adherence, integrated out via Gauss--Hermite quadrature. For our two-state specification, the probability of being in the high-adherence state at round $t$ given origin state $j$ at round $t-1$ is modeled as a logistic function of the transition propensity:
\begin{equation}\label{eq:nhmm_transition}
P(S_{it} = 0 \mid S_{i,t-1} = j) = \sigma\!\left(\eta_{i,j,t} + c_{j,0}\right),
\end{equation}
where $\sigma(\cdot)$ is the logistic function and $c_{j,0}$ is an origin-state-specific transition threshold that captures the baseline propensity to remain in (or transition to) the high-adherence state, absent any covariate effects. A positive element of $\boldsymbol{\beta}_j$ pushes probability mass toward the high-adherence state in the next round, while a negative element pushes it away. Equation~\eqref{eq:nhmm_transition} is the two-state case of the general cumulative ordered logit transition model stated in Appendix~\ref{sec:nhmm_appendix}.

We estimate the model by maximizing the marginal likelihood (forward algorithm with Gauss--Hermite integration) on the treatment-group subsample (147 participants, 3{,}862 participant-round observations), separately for each gender composition $g \in \{\text{female-only, male-only, mixed-gender}\}$. Full technical details are provided in Appendix~\ref{sec:nhmm_appendix}. We focus on the lagged-profit coefficient as the key mechanism parameter because profit is the participant's incentive-compatible objective and the most direct signal of whether their pricing strategy is succeeding. By comparing this coefficient across compositions, we assess whether higher profit predicts stronger (positive coefficient) or weaker (negative coefficient) AI adherence within each gender composition.


\section{Results}\label{sec:results}

Section~\ref{sec:ai_char} first characterizes the AI's recommendations in isolation as a baseline check. Section~\ref{sec:rq1} then estimates the average effect of AI pricing recommendations on market prices and profits (RQ1). Section~\ref{sec:rq2} tests whether and how gender composition moderates this effect (RQ2). Section~\ref{sec:rq3} examines the behavioral mechanism through which gender composition may produce heterogeneous AI effects, using the Non-Homogeneous Hidden Markov Model specified in Section~\ref{sec:econometric} (RQ3).
\subsection{AI Recommendation Characteristics}\label{sec:ai_char}

We first characterize the AI's pricing behavior in isolation, since its recommendations are the input signal participants act on in the treatment condition. In this baseline simulation, three independent GPT-5 mini agents play the Bertrand game against each other for 300 rounds with no human involvement.

Each agent's prompt is identical to the one used to generate AI recommendations in the treatment condition (full text in Listing~\ref{lst:llm_prompt} ). It supplies per-round market information matching what human participants see: own cost, the price ceiling, cluster size, the previous round's own outcomes (price, market share, profit, and cumulative profit), and the previous round's market aggregates (minimum, average, and maximum prices). The prompt instructs the model to ``set prices that maximize your own profit in the long run'' and to ``explore many different pricing strategies,'' and feeds back the model's previous-round strategy as a form of persistent memory across rounds. Critically, the prompt contains no information about Nash or monopoly benchmarks and no language encouraging or discouraging coordination.
Table~\ref{tab:ai_char} summarizes the resulting price distribution and convergence over the experiment-relevant window (rounds 1--30).

\begin{table}[!htbp]
\centering
\caption{AI Price Recommendation Characteristics (LLM-Only Benchmark)}
\label{tab:ai_char}
\begin{threeparttable}
\begin{tabular}{lc}
\toprule
\multicolumn{2}{l}{Panel A: Distribution of AI Prices (Rounds 1--30)} \\[2pt]
Mean & \$2.08 \\
Standard deviation & \$0.18 \\
Minimum & \$1.84 \\
Maximum & \$2.60 \\
\% above Nash equilibrium (\$1.37) & 100.0\% \\[6pt]
\multicolumn{2}{l}{Panel B: Convergence Over Rounds} \\[2pt]
Rounds 1--10: mean price (SD) & \$2.27 (0.17) \\
Rounds 11--20: mean price (SD) & \$2.03 (0.11) \\
Rounds 21--30: mean price (SD) & \$1.95 (0.06) \\[2pt]
Mean within-round range across 3 sellers & \$0.18 \\
\bottomrule
\end{tabular}
\begin{tablenotes}
\footnotesize
\item Notes. Three independent GPT-5 mini agents play the same Bertrand pricing game for 300 rounds with no human involvement. Panel~A reports the distribution of all 90 seller-round AI price recommendations in the experiment-relevant window (rounds 1--30). Panel~B shows convergence by dividing the window into three 10-round windows. The Nash equilibrium price is \$1.37; the monopoly price is \$2.00.
\end{tablenotes}
\end{threeparttable}
\end{table}

Two patterns stand out. First, the AI systematically recommends supra-competitive prices: every recommendation in the experiment-relevant window exceeds the Nash equilibrium of \$1.37, with a mean of \$2.08 (SD = \$0.18), roughly 52\% above the Nash benchmark. Since the prompt contains no information about Nash or monopoly prices and no language encouraging coordination, this pricing elevation is emergent rather than instructed. Inspection of the LLM's chain-of-thought rationales reveals that it independently discovers concepts such as ``tacit cooperation,'' ``tit-for-tat punishment,'' and ``supra-competitive pricing,'' drawing on strategic reasoning internalized during pre-training.

Second, the AI's recommendations are tightly aligned across sellers within a market and stabilize over the experiment-relevant window. The mean within-round price range across the three sellers is only \$0.18. The temporal spread of recommendations also shrinks over the same horizon, with the standard deviation falling from \$0.17 in rounds 1--10 to \$0.06 in rounds 21--30. Taken together, the two patterns position the AI as a coordination-facilitating anchor that delivers similar supra-competitive price signals to every seller in a cluster.

In the treatment condition, where the AI advises human sellers who observe market feedback and adjust their own prices, the supra-competitive tendency persists on aggregate but varies across compositions. Table~\ref{tab:ai_char_treatment} reports the pooled distribution: the AI recommends prices above the Nash equilibrium in 84.7\% of all seller-rounds (mean = \$1.84, a 34\% premium over Nash), and in 73.2\% of cluster-rounds all three sellers simultaneously receive recommendations exceeding Nash. Within-round cross-seller alignment remains moderate (mean range = \$0.32), tightening from \$0.60 in rounds 1--10 to \$0.27 in rounds 21--30 as the AI's adaptive algorithm converges on market conditions.

However, because the AI observes each cluster's pricing history and adapts, these pooled statistics mask important differences across gender compositions. Table~\ref{tab:ai_char_by_comp} disaggregates by composition. In early rounds (1--10), before adherence patterns have fully formed, all three compositions receive supra-competitive recommendations (female-only: \$2.28; male-only: \$1.94; mixed: \$2.06). Over subsequent rounds, the recommendations diverge: they remain elevated in female-only clusters (\$1.84 in rounds 21--30, above those sellers' \$1.55 no-AI baseline); converge in male-only clusters to \$1.80, close to the \$1.81 no-AI baseline; and decline sharply in mixed-gender clusters (\$1.42) as the AI adapts to those sellers' lower prices. This adaptive divergence means the AI does not provide a uniform signal across compositions. We return to this when interpreting the heterogeneous treatment effects in \S\ref{sec:rq2}.

\begin{table}[!htbp]
\centering
\caption{AI Price Recommendation Characteristics (Treatment Condition)}
\label{tab:ai_char_treatment}
\begin{threeparttable}
\begin{tabular}{lc}
\toprule
\multicolumn{2}{l}{Panel A: Distribution of AI Prices (Rounds 1--30)} \\[2pt]
Mean & \$1.84 \\
Standard deviation & \$0.48 \\
Minimum & \$1.00 \\
Maximum & \$4.20 \\
\% above Nash equilibrium (\$1.37) & 84.7\% \\
\% of cluster-rounds with all 3 recs $>$ Nash & 73.2\% \\[6pt]
\multicolumn{2}{l}{Panel B: Convergence Over Rounds} \\[2pt]
Rounds 1--10: mean price (SD) & \$2.09 (0.60) \\
Rounds 11--20: mean price (SD) & \$1.70 (0.33) \\
Rounds 21--30: mean price (SD) & \$1.67 (0.27) \\[2pt]
Mean within-round range across 3 sellers & \$0.32 \\
\bottomrule
\end{tabular}
\begin{tablenotes}
\footnotesize
\item Notes. AI recommendations delivered to human sellers in treatment-group clusters ($N = 3{,}937$ seller-rounds across 49 clusters). Panel~A reports the distribution of AI price recommendations over all 30 rounds. Panel~B shows convergence by phase. The Nash equilibrium price is \$1.37; the monopoly price is \$2.00. Unlike the LLM-only benchmark (Table~\ref{tab:ai_char}), the AI here observes human sellers' actual pricing decisions and adapts its recommendations accordingly. Table~\ref{tab:ai_char_by_comp} disaggregates these statistics by gender composition.
\end{tablenotes}
\end{threeparttable}
\end{table}

\begin{table}[!htbp]
\centering
\caption{AI Price Recommendation Characteristics by Gender Composition (Treatment Condition)}
\label{tab:ai_char_by_comp}
\begin{threeparttable}
\begin{tabular}{lccc}
\toprule
& \textbf{Female-only} & \textbf{Male-only} & \textbf{Mixed-gender} \\
\midrule
\multicolumn{4}{l}{Panel A: Distribution of AI Prices (Rounds 1--30)} \\[2pt]
Mean & \$2.03 & \$1.84 & \$1.69 \\
Standard deviation & \$0.49 & \$0.33 & \$0.52 \\
Minimum & \$1.00 & \$1.00 & \$1.00 \\
Maximum & \$4.00 & \$4.00 & \$4.20 \\
\% above Nash (\$1.37) & 90.4\% & 93.2\% & 73.2\% \\
Seller-rounds & 1{,}162 & 1{,}260 & 1{,}515 \\[6pt]
\multicolumn{4}{l}{Panel B: Convergence Over Rounds} \\[2pt]
Rounds 1--10: mean (SD) & \$2.28 (0.59) & \$1.94 (0.47) & \$2.06 (0.66) \\
Rounds 11--20: mean (SD) & \$1.90 (0.36) & \$1.76 (0.23) & \$1.50 (0.23) \\
Rounds 21--30: mean (SD) & \$1.84 (0.25) & \$1.80 (0.18) & \$1.42 (0.16) \\[2pt]
Mean within-round range & \$0.32 & \$0.22 & \$0.40 \\
\bottomrule
\end{tabular}
\begin{tablenotes}
\footnotesize
\item Notes. Same treatment-condition sample as Table~\ref{tab:ai_char_treatment}, disaggregated by cluster gender composition. Because the AI observes each cluster's pricing history and adapts, recommendation levels diverge across compositions over rounds. The Nash equilibrium price is \$1.37; the monopoly price is \$2.00.
\end{tablenotes}
\end{threeparttable}
\end{table}

\subsection{Pooled Effect of AI Pricing Recommendations}\label{sec:rq1}

We first examine the pooled effect of AI pricing recommendations on cluster-round market outcomes. Specifically, we use cluster average price and aggregate cluster profit per round as the dependent variables. Table~\ref{tab:summary_stats} reports raw summary statistics by treatment condition, and Table~\ref{tab:pooled} reports the estimation results of specification~\eqref{eq:pooled}, with cluster average price in Column~(1) and aggregate cluster profit in Column~(2). All specifications include round fixed effects and cluster-level controls, with standard errors clustered at the cluster level.

\begin{table}[!htbp]
\centering
\caption{Summary Statistics by Treatment Condition}
\label{tab:summary_stats}
\begin{threeparttable}
\begin{tabular}{lcccc}
\toprule
& \multicolumn{2}{c}{\textbf{Control}} & \multicolumn{2}{c}{\textbf{Treatment}} \\
\cmidrule(lr){2-3} \cmidrule(lr){4-5}
& Mean & SD & Mean & SD \\
\midrule
Average Price (\$) & 1.64 & 0.36 & 1.79 & 0.47 \\
Aggregate Profit (\$) & 0.45 & 0.28 & 0.47 & 0.23 \\
\% of Cluster-Rounds Above Nash & 80.1 & --- & 81.0 & --- \\
\midrule
Cluster-Rounds & \multicolumn{2}{c}{1{,}175} & \multicolumn{2}{c}{1{,}255} \\
Clusters & \multicolumn{2}{c}{42} & \multicolumn{2}{c}{49} \\
\bottomrule
\end{tabular}
\begin{tablenotes}
\footnotesize
\item Notes. Sample restricted to cluster-rounds in which all three sellers have valid price and profit data ($N = 2{,}430$). Average Price is the mean of the three sellers' prices per cluster-round. Aggregate Profit is the sum of their per-round profits. The Nash equilibrium price is \$1.37.
\end{tablenotes}
\end{threeparttable}
\end{table}

\begin{table}[!htbp]
\centering
\caption{Pooled Panel Regression Results}
\label{tab:pooled}
\begin{threeparttable}
\begin{tabular}{lcc}
\toprule
& \textbf{(1)} & \textbf{(2)} \\
& \textbf{Average Price} & \textbf{Aggregate Profit} \\
\midrule
Treatment & $0.1443^{***}$ & $0.0360$ \\
& $(0.0494)$ & $(0.0372)$ \\[2pt]
FemaleOnly & $-0.0960^{**}$ & $-0.0503$ \\
& $(0.0458)$ & $(0.0472)$ \\[2pt]
Mixed & $-0.2755^{***}$ & $-0.1946^{***}$ \\
& $(0.0578)$ & $(0.0537)$ \\[2pt]
Average Age & $-0.0133$ & $-0.0120^{*}$ \\
& $(0.0108)$ & $(0.0072)$ \\[2pt]
Age Diversity & $-0.0102$ & $-0.0039$ \\
& $(0.0101)$ & $(0.0062)$ \\[2pt]
Ethnicity/Race Diversity & $-0.0975$ & $0.0302$ \\
& $(0.0922)$ & $(0.0975)$ \\[2pt]
Major Diversity & $-0.1212$ & $0.2551^{***}$ \\
& $(0.0956)$ & $(0.0581)$ \\[2pt]
E-Commerce Prevalence & $0.0350$ & $0.0155$ \\
& $(0.0892)$ & $(0.0802)$ \\[2pt]
Prior GPT Trust & $0.0310$ & $0.1120^{*}$ \\
& $(0.0620)$ & $(0.0675)$ \\[2pt]
\midrule
Observations & 2{,}430 & 2{,}430 \\
Clusters & 91 & 91 \\
$R^2$ & 0.165 & 0.231 \\
Round FE & Yes & Yes \\
\bottomrule
\end{tabular}
\begin{tablenotes}
\footnotesize
\item Notes. Sample restricted to cluster-rounds in which all three sellers have valid price and profit data. Standard errors clustered at the cluster level in parentheses. $^{*}\,p<0.10$; $^{**}\,p<0.05$; $^{***}\,p<0.01$. Round fixed effects included. $R^2$ is from the time-demeaned regression. Male-only clusters are the reference category for gender composition dummies.
\end{tablenotes}
\end{threeparttable}
\end{table}

Across all gender compositions, the presence of AI pricing recommendations raises average price by approximately 14.4 cents per round, an 8.8\% increase relative to the control mean of \$1.64 ($\hat{\beta}_1 = 0.144$, $p < 0.01$). The pooled effect on aggregate cluster profit is positive but does not reach significance ($\hat{\beta}_1 = 0.036$, $p = 0.334$). As we show in §\ref{sec:rq2}, this reflects substantial heterogeneity: the profit effect is concentrated in female-only clusters, where it is large and significant, while the absence of an effect in male-only and mixed-gender clusters dilutes the pooled estimate.

In the average-price regression (Column~1), the coefficients on the female-only and mixed-gender dummies are also negative and statistically significant relative to the male-only baseline.

\medskip
\noindent\textbf{Finding 1.} AI pricing recommendations significantly increase cluster average price, consistent with AI recommendations nudging participants toward supra-competitive pricing. The pooled profit effect is positive but not significant, reflecting heterogeneity across compositions documented in §\ref{sec:rq2}.

\subsection{Effect of AI Pricing Recommendations by Gender Composition}\label{sec:rq2}

Having documented the pooled effects of AI on prices and profits, we next examine whether these aggregate estimates mask heterogeneity across gender compositions. Absent AI pricing recommendations, male-only control clusters price the highest at \$1.81 per round, followed by female-only (\$1.55) and mixed-gender (\$1.51) clusters. All three compositions sustain prices above the Nash equilibrium of \$1.37, with male-only clusters pricing significantly higher than both female-only and mixed-gender clusters ($p < 0.001$ in both pairwise comparisons). To examine whether the effect of AI pricing recommendations varies across these gender compositions, we estimate Equation~\eqref{eq:subsample} separately within each subsample. Tables~\ref{tab:price_by_gender} and \ref{tab:profit_by_gender} report the results on cluster average price and aggregate cluster profit, respectively.

\begin{table}[!htbp]
\centering
\caption{Panel Regression Results: Average Price by Gender Composition}
\label{tab:price_by_gender}
\begin{threeparttable}
\begin{tabular}{lccc}
\toprule
& \textbf{(1)} & \textbf{(2)} & \textbf{(3)} \\
& \textbf{Female-Only} & \textbf{Male-Only} & \textbf{Mixed-Gender} \\
\midrule
Treatment & $0.4465^{***}$ & $-0.0262$ & $0.0200$ \\
& $(0.0551)$ & $(0.0376)$ & $(0.0746)$ \\[2pt]
Female Proportion & -- & -- & $0.0738$ \\
& & & $(0.2426)$ \\[2pt]
Average Age & $-0.0104$ & $-0.0395^{**}$ & $0.0122$ \\
& $(0.0178)$ & $(0.0197)$ & $(0.0109)$ \\[2pt]
Age Diversity & $-0.0165$ & $-0.0413$ & $0.0034$ \\
& $(0.0385)$ & $(0.0325)$ & $(0.0101)$ \\[2pt]
Ethnicity/Race Diversity & $-0.0172$ & $-0.0975$ & $0.2363^{*}$ \\
& $(0.1260)$ & $(0.0737)$ & $(0.1378)$ \\[2pt]
Major Diversity & $-0.1050$ & $-0.0627$ & $-0.1785$ \\
& $(0.1475)$ & $(0.1042)$ & $(0.1525)$ \\[2pt]
E-Commerce Prevalence & $0.1078$ & $0.0594$ & $0.1399$ \\
& $(0.1286)$ & $(0.0779)$ & $(0.1645)$ \\[2pt]
Prior GPT Trust & $-0.1550^{**}$ & $0.0528$ & $0.2234^{***}$ \\
& $(0.0777)$ & $(0.0650)$ & $(0.0774)$ \\[2pt]
\midrule
Observations & 793 & 841 & 796 \\
Clusters & 30 & 30 & 31 \\
$R^2$ & 0.345 & 0.073 & 0.094 \\
Round FE & Yes & Yes & Yes \\
\bottomrule
\end{tabular}
\begin{tablenotes}
\footnotesize
\item Notes. Sample restricted to cluster-rounds in which all three sellers have valid price and profit data. Standard errors clustered at the cluster level in parentheses. $^{*}\,p<0.10$; $^{**}\,p<0.05$; $^{***}\,p<0.01$. Round fixed effects included in all specifications. $R^2$ is from the time-demeaned regression.
\end{tablenotes}
\end{threeparttable}
\end{table}

\begin{table}[!htbp]
\centering
\caption{Panel Regression Results: Aggregate Profit by Gender Composition}
\label{tab:profit_by_gender}
\begin{threeparttable}
\begin{tabular}{lccc}
\toprule
& \textbf{(1)} & \textbf{(2)} & \textbf{(3)} \\
& \textbf{Female-Only} & \textbf{Male-Only} & \textbf{Mixed-Gender} \\
\midrule
Treatment & $0.1726^{***}$ & $0.0346$ & $-0.0200$ \\
& $(0.0314)$ & $(0.0600)$ & $(0.0593)$ \\[2pt]
Female Proportion & -- & -- & $0.0915$ \\
& & & $(0.1985)$ \\[2pt]
Average Age & $-0.0106$ & $-0.0615^{***}$ & $0.0109$ \\
& $(0.0111)$ & $(0.0180)$ & $(0.0076)$ \\[2pt]
Age Diversity & $0.0427$ & $0.0659^{*}$ & $0.0057$ \\
& $(0.0270)$ & $(0.0365)$ & $(0.0069)$ \\[2pt]
Ethnicity/Race Diversity & $-0.1310$ & $0.0116$ & $0.1340$ \\
& $(0.0951)$ & $(0.1551)$ & $(0.1062)$ \\[2pt]
Major Diversity & $0.2858^{***}$ & $0.2826^{**}$ & $-0.0878$ \\
& $(0.0756)$ & $(0.1165)$ & $(0.0950)$ \\[2pt]
E-Commerce Prevalence & $-0.0120$ & $-0.0718$ & $0.0140$ \\
& $(0.0720)$ & $(0.1418)$ & $(0.1030)$ \\[2pt]
Prior GPT Trust & $-0.0270$ & $0.1669$ & $0.0838$ \\
& $(0.0340)$ & $(0.1344)$ & $(0.0614)$ \\[2pt]
\midrule
Observations & 793 & 841 & 796 \\
Clusters & 30 & 30 & 31 \\
$R^2$ & 0.334 & 0.256 & 0.048 \\
Round FE & Yes & Yes & Yes \\
\bottomrule
\end{tabular}
\begin{tablenotes}
\footnotesize
\item Notes. Sample restricted to cluster-rounds in which all three sellers have valid price and profit data. Standard errors clustered at the cluster level in parentheses. $^{*}\,p<0.10$; $^{**}\,p<0.05$; $^{***}\,p<0.01$. Round fixed effects included in all specifications. $R^2$ is from the time-demeaned regression.
\end{tablenotes}
\end{threeparttable}
\end{table}

In female-only clusters, the presence of AI pricing recommendations raises cluster average price by approximately 45 cents per round ($\hat{\beta}_1 = 0.447$, $p < 0.001$), a 29\% increase relative to the female-only control mean of \$1.55. This pricing increase translates to approximately 17 cents per round in aggregate cluster profit ($\hat{\beta}_1 = 0.173$, $p < 0.001$), a 39\% improvement. In contrast, in male-only clusters, the treatment effect on price is small and not statistically significant ($\hat{\beta}_1 = -0.026$, $p = 0.486$), and the profit effect is likewise insignificant ($\hat{\beta}_1 = 0.035$, $p = 0.565$). Mixed-gender clusters show no significant effect on either outcome ($\hat{\beta}_1 = 0.020$, $p = 0.788$ for price; $\hat{\beta}_1 = -0.020$, $p = 0.736$ for profit).

The by-composition results show that the significant pooled treatment effect is driven by female-only clusters, where the AI recommendation acts as a coordination focal point that pulls seller prices toward the supra-competitive range. In male-only and mixed-gender clusters, AI pricing recommendations produce no significant response. Appendix~\ref{sec:interaction} confirms this heterogeneity formally: a pooled model with Treatment $\times$ Composition interactions yields a large and significant interaction for female-only clusters on price ($p < 0.001$), while the interaction for mixed-gender clusters is insignificant.

Part of this heterogeneity reflects differences in the anchoring gap between AI recommendations and no-AI baseline prices. As documented in Table~\ref{tab:ai_char_by_comp}, the AI's adaptive recommendations diverge across compositions. Female-only control clusters price at \$1.55, well below the AI's early-round recommendations (\$2.28 in rounds 1--10), leaving substantial room for the AI to pull prices upward. Male-only control clusters, by contrast, already price near the monopoly benchmark at \$1.81; the AI's early-round recommendations exceed this baseline only modestly (\$1.94 versus \$2.28 for female-only) and then converge to \$1.80 by rounds 21--30, leaving little room for an additional effect. However, anchoring distance alone cannot fully account for the pattern: mixed-gender clusters have a baseline (\$1.51) nearly as low as female-only and receive similar early-round recommendations (\$2.06), yet the treatment effect is transient and statistically insignificant. This points to a second factor beyond anchoring distance: how sellers respond to the outcomes that follow. We examine this mechanism in \S\ref{sec:rq3}.

\begin{figure}[!htbp]
\centering
\includegraphics[width=\textwidth]{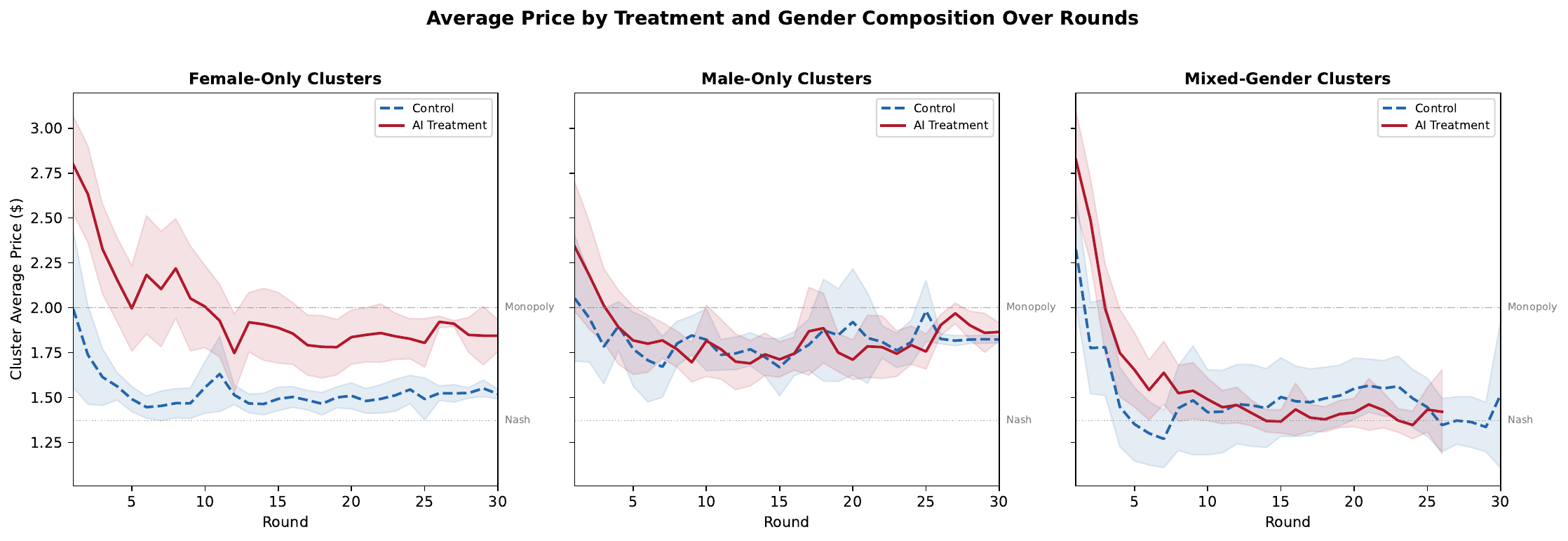}
\caption{Average Price by Treatment and Gender Composition Over Rounds}
\label{fig:round_dynamics}
\begin{minipage}{\textwidth}
\footnotesize
Notes. Each panel plots the cluster-level average price per round for treatment (solid) and control (dashed) groups within a gender composition. Shaded bands denote 95\% confidence intervals based on cross-cluster standard errors of the mean. Horizontal reference lines mark the Nash equilibrium price (\$1.37) and the monopoly price (\$2.00).
\end{minipage}
\end{figure}

To examine whether these by-composition effects of AI pricing recommendations persist over time, we plot cluster average prices by treatment arm and gender composition across rounds in Figure~\ref{fig:round_dynamics}. In female-only markets, the treatment effect peaks in the first 10 rounds at $+\$0.54$ ($p < 0.001$) and remains positive and significant in rounds 11--20 ($+\$0.41$, $p < 0.001$) and 21--30 ($+\$0.37$, $p < 0.001$). In male-only markets, the treatment effect is negligible across all phases, with no significant time trend. In mixed-gender markets, the treatment effect is positive but not significant in rounds 1--10 ($+\$0.16$, $p = 0.199$) and vanishes by rounds 11--20. These dynamics confirm that AI pricing recommendations persistently raise prices in female-only markets, have no effect in male-only markets, and produce only a transient effect in mixed-gender markets.

Three robustness checks probe these results. The first two target different failure modes of the same broad concern: treatment is assigned at the session level with only seven sessions. The first is inferential: cluster-robust standard errors may overstate precision if unobserved shocks are correlated within a session. Appendix~\ref{sec:session_clustering} therefore re-estimates each specification under three inference procedures that account for the session-level treatment assignment: session-clustered standard errors with small-sample $t$ critical values, the wild cluster bootstrap (WCB) of \citet{cameron2008bootstrap}, and Fisher randomization inference (RI). Both female-only effects are strongly significant under session-robust inference ($p = 0.0009$ for price and $p = 0.0006$ for profit). The WCB and RI $p$-values are near conventional thresholds (price: WCB $0.047$, RI $0.114$; profit: WCB $0.039$, RI $0.114$), reflecting the limited resolution of $2^7 = 128$ bootstrap draws and $35$ permutations rather than a weak underlying signal: the session-clustered $t$-statistics are $6.1$ (price) and $6.6$ (profit). All male-only and mixed-gender WCB/RI $p$-values exceed $0.31$, consistent with the absence of a treatment effect in these cells, though the session-level MDE analysis below shows that these cells are underpowered, so smaller effects cannot be ruled out.

The second is influence: even if inference treats sessions as the assignment units, the female-only point estimate could still be driven by one anomalous treated or control session. Appendix~\ref{sec:session_clustering} therefore reports a leave-one-session-out check for female-only markets. The female-only price effect remains positive and significant under every session omission, ranging from $0.324$ to $0.530$; the profit effect is also positive throughout, ranging from $0.151$ to $0.279$. This pattern indicates that the female-only result is not mechanically driven by a single session.

The third concern is power: null effects in male-only and mixed-gender markets could reflect insufficient sensitivity rather than true absence of economically meaningful effects. Following \citet{baroudi1989problem}, Table~\ref{tab:mde} reports the minimum detectable effect (MDE) at 80\% power for all three compositions. The MDE is the smallest true treatment effect that the study would detect as statistically significant (at the 5\% level) at least 80\% of the time. The female-only cells are well-powered, with observed effects exceeding the detection threshold (ratios of 2.77 for price and 1.88 for profit). For price, the MDEs in male-only and mixed-gender cells are well below the female-only point estimate (e.g., the male-only price MDE of 0.110 is far below the female-only price effect of 0.447), meaning the study is well-powered to detect a female-only-magnitude price effect in these cells; the near-zero observed effects therefore credibly rule out price treatment effects as large as those in female-only clusters. For profit, the male-only MDE (0.176) slightly exceeds the female-only profit estimate (0.173), so the study cannot rule out a female-sized profit effect in male-only clusters at 80\% power. The ratios well below one for male-only and mixed-gender cells indicate that the observed effects fall short of their own detection thresholds, so we cannot rule out smaller effects below the MDE in these compositions. Because treatment is randomized at the session level, Appendix~\ref{sec:mde_session} recomputes MDEs using session-robust standard errors and $t(G{-}1)$ critical values; the female-only effects remain well-powered (ratios of 1.83 and 1.97).

\begin{table}[!htbp]
\centering
\caption{Minimum Detectable Effect and Statistical Power by Gender Composition}
\label{tab:mde}
\begin{threeparttable}
\begin{tabular}{llccccc}
\toprule
\textbf{Outcome} & \textbf{Composition} & \textbf{Clusters} & $\hat{\beta}$ & \textbf{SE} & \textbf{MDE} & \textbf{Ratio} \\
\midrule
\multirow{3}{*}{Average Price}
& Female-Only   & 30 & $0.447$ & $0.055$ & $0.162$ & $2.77$ \\
& Male-Only     & 30 & $-0.026$ & $0.038$ & $0.110$ & $0.24$ \\
& Mixed-Gender  & 31 & $0.020$ & $0.075$ & $0.219$ & $0.09$ \\[2pt]
\multirow{3}{*}{Aggregate Profit}
& Female-Only   & 30 & $0.173$ & $0.031$ & $0.092$ & $1.88$ \\
& Male-Only     & 30 & $0.035$ & $0.060$ & $0.176$ & $0.20$ \\
& Mixed-Gender  & 31 & $-0.020$ & $0.059$ & $0.174$ & $0.12$ \\
\bottomrule
\end{tabular}
\begin{tablenotes}
\footnotesize
\item Notes. MDE is the minimum detectable effect at 80\% power and 5\% significance, computed as $\text{SE} \times (t_{\alpha/2,\,df} + t_{\beta,\,df})$ where SE is the cluster-robust standard error of the treatment coefficient. Ratio $= |\hat{\beta}| / \text{MDE}$; values above 1 indicate the observed effect exceeds the detection threshold.
\end{tablenotes}
\end{threeparttable}
\end{table}

\medskip
\noindent\textbf{Finding 2.} AI pricing recommendations produce divergent market outcomes across gender compositions. They significantly raise prices and profits in female-only clusters but have no significant effect in male-only or mixed-gender clusters.

\subsection{Behavioral Mechanism: AI Adherence Dynamics}\label{sec:rq3}

The by-composition results in §\ref{sec:rq2} describe what happens but not the dynamic through which gender composition produces such divergent responses to AI pricing recommendations. To characterize the behavioral mechanism, we estimate the NHMM specified in §\ref{sec:econometric} separately for each composition on the treatment-group subsample (147 participants, 3{,}862 participant-rounds; full technical details in Appendix~\ref{sec:nhmm_appendix}). Table~\ref{tab:nhmm} reports the estimation results in three panels: the distribution of price deviations by state (Panel~A), the effect of lagged profit on state transitions (Panel~B), and the implied probability of remaining adherent at different profit levels (Panel~C).

\begin{table}[!htbp]
\centering
\caption{NHMM Estimation Results by Gender Composition}
\label{tab:nhmm}
\begin{threeparttable}
\begin{tabular}{lccc}
\toprule
& \textbf{Female-Only} & \textbf{Male-Only} & \textbf{Mixed-Gender} \\
\midrule
\multicolumn{4}{l}{Panel A: Emission Distributions} \\[2pt]
State 0 mean ($\mu_0$) & $0.018^{***}$ & $0.017^{***}$ & $0.119^{***}$ \\
& $(0.001)$ & $(0.001)$ & $(0.004)$ \\[2pt]
State 0 SD ($\sigma_0$) & $0.019$ & $0.015$ & $0.108$ \\[2pt]
State 1 mean ($\mu_1$) & $0.415^{***}$ & $0.384^{***}$ & $0.723^{***}$ \\
& $(0.022)$ & $(0.021)$ & $(0.037)$ \\[2pt]
State 1 SD ($\sigma_1$) & $0.508$ & $0.464$ & $0.609$ \\[2pt]
Random effect SD ($\sigma_\xi$) & $0.597$ & $0.603$ & $0.705$ \\[6pt]

\multicolumn{4}{l}{Panel B: Effect of Lagged Profit on Transition Toward Adherent State} \\[2pt]
Adherent $\to$ Adherent ($\beta_0$) & $0.642^{**}$ & $-7.216^{***}$ & $-1.282^{**}$ \\
& $(0.270)$ & $(1.212)$ & $(0.525)$ \\[2pt]
Non-adherent $\to$ Adherent ($\beta_1$) & $0.056$ & $1.981^{***}$ & $0.211$ \\
& $(0.167)$ & $(0.399)$ & $(0.170)$ \\[6pt]

\multicolumn{4}{l}{Panel C: Implied Adherence Persistence $P(\text{stay in State 0})$} \\[2pt]
At mean covariates & $0.692$ & $0.987$ & $0.946$ \\[2pt]
At $+1$ SD lagged profit & $0.810$ & $0.052$ & $0.829$ \\[2pt]
At $-1$ SD lagged profit & $0.541$ & $1.000$ & $0.984$ \\[6pt]

\midrule
Log-likelihood & $550.3$ & $1{,}356.7$ & $307.0$ \\
BIC & $-931.5$ & $-2{,}542.2$ & $-439.0$ \\
Participants & 45 & 45 & 57 \\
Observations & 1{,}150 & 1{,}251 & 1{,}461 \\
\bottomrule
\end{tabular}
\begin{tablenotes}
\footnotesize
\item Notes. Two-state NHMM estimated separately by gender composition on treatment-group participants only (147 participants, 3{,}862 participant-rounds). Standard errors from the numerical Hessian in parentheses. $^{*}\,p<0.10$; $^{**}\,p<0.05$; $^{***}\,p<0.01$. Panel~B reports the coefficient on standardized lagged profit in the transition model~\eqref{eq:nhmm_transition}; positive (negative) values indicate that higher profit pushes probability mass toward (away from) the adherent state. Panel~C computes $P(\text{stay adherent}) = \sigma(c_{0,0} + \beta_0 \cdot \Delta)$ for $\Delta \in \{0, +1, -1\}$ SD of lagged profit, with all other covariates at their means and $\xi_i = 0$. Each model has 24 free parameters estimated by BFGS with 15 random initializations and 15-point Gauss--Hermite quadrature.
\end{tablenotes}
\end{threeparttable}
\end{table}

Panel~A shows that all three gender compositions exhibit a two-state structure: a high-adherence state and a low-adherence state. In female-only and male-only clusters, the high-adherence state involves near-exact compliance with the AI's recommendation ($\mu_0 \approx 0.018$), while the low-adherence state features moderate deviations ($\mu_1 \approx 0.40$). Mixed-gender clusters differ: even the high-adherence state exhibits substantial deviation from the AI's recommendation ($\mu_0 = 0.119$), roughly seven times larger than in single-gender clusters, and the non-adherent state features large deviations ($\mu_1 = 0.723$). This suggests that the behavioral heterogeneity inherent in mixed-gender groups prevents the tight price alignment that characterizes single-gender adherence.

Panel~B shows that sellers in female-only and male-only markets respond in opposite ways to earning a higher profit after following the AI's pricing recommendation in the previous round. In female-only markets, the coefficient is positive ($\hat{\beta}_0 = 0.64$, $p = 0.018$), indicating that sellers become more likely to follow the AI again after earning a higher profit. By contrast, in male-only markets, the coefficient is strongly negative ($\hat{\beta}_0 = -7.22$, $p < 0.001$), indicating that sellers become less likely to follow the AI again after earning a higher profit.

Panel~C shows that these opposing responses are substantial in magnitude. In female-only markets, when the previous round's realized profit is high (i.e., one standard deviation above average profit), the probability of continuing to follow the AI's pricing recommendation rises from 69\% to 81\%, a 12-percentage-point increase. In male-only markets, the same profit increase is associated with a drop in this probability from 99\% to 5\%, a 94-percentage-point collapse.

These dynamics are consistent with opposite feedback patterns that help resolve the puzzle of divergent market outcomes we observe in §\ref{sec:rq2}. In female-only clusters, sellers follow the AI's supra-competitive recommendations and earn high profits. Higher profits predict continued adherence in subsequent rounds, consistent with prices converging toward the AI's price anchor and coordination being sustained. This positive dynamic association is consistent with the significant positive AI treatment effect on both prices and profits in female-only clusters. However, in male-only clusters, higher profits predict lower adherence ($\hat{\beta}_0 = -7.22$), with sellers tending to deviate from the AI's recommendation. Although profitable deviation predicts a return toward the AI ($\hat{\beta}_1 = 1.98$), the net association is consistent with coordination failing to sustain in male-only markets. Combined with the narrow gap between male-only control prices (\$1.81) and the AI's adaptive recommendations (\$1.80 by rounds 21--30), the AI does not raise male-only market prices above their already-high no-AI baseline. This contrast between the two dynamic associations, positive in female-only markets and negative in male-only markets, is consistent with the divergent outcomes we observe between single-gender compositions.

Mixed-gender clusters present a third pattern. Sellers in these clusters follow the AI's recommendations only loosely from the start. Even in the high-adherence state, they deviate from the AI by roughly \$0.12 on average ($\mu_0 = 0.119$). This loose alignment is just enough to pull prices modestly toward the AI's supra-competitive price anchor in early rounds, producing a small positive treatment effect ($+\$0.24$ in rounds 1--10). However, higher profits predict weaker adherence even in this loose-adherence state ($\hat{\beta}_0 = -1.28$, $p = 0.015$), with the probability of continuing to follow the AI falling from 95\% to 83\% after a high-profit round. This negative association is consistent with the partial coordination collapsing and the treatment effect gradually disappearing in later rounds.

Appendix~\ref{sec:direct_adherence} confirms these patterns using a direct dynamic adherence model that does not rely on latent states or parametric emission distributions: a linear probability model of binary adherence on lagged adherence, lagged profit, and their interaction. The interaction coefficient is significantly positive for female-only clusters and significantly negative for male-only clusters, matching the NHMM transition estimates.

\medskip
\noindent\textbf{Finding 3.} Sellers in female-only and male-only markets exhibit opposite associations between profit and subsequent adherence to the AI's price recommendation. Higher profit predicts stronger adherence in female-only markets, consistent with a coordination-sustaining dynamic. By contrast, higher profit predicts weaker adherence in male-only markets, consistent with a coordination-breaking dynamic. Because the NHMM is estimated separately by market composition, these are composition-level effects; the design does not isolate individual gender independently of the strategic environment.


\section{Discussion}\label{sec:discussion}

AI pricing recommendations produce divergent market outcomes across gender compositions. They significantly raise prices and profits in female-only clusters but have no significant effect in male-only or mixed-gender clusters. This divergence is linked to opposite dynamic associations between profit and subsequent adherence in female-only and male-only markets. Higher profit predicts stronger adherence in female-only markets, consistent with a coordination-sustaining dynamic, but weaker adherence in male-only markets, consistent with a coordination-breaking one.
\subsection{Behavioral Mechanism}\label{sec:mechanism}

Two interpretive limitations apply to the NHMM evidence. First, the model is estimated separately by market composition, not individual gender: because every seller in a single-gender market shares that market's gender, the price data alone cannot separate an individual-gender effect from an effect of the single-gender environment itself. Second, the NHMM is estimated on the treatment group only, where sellers face both profit feedback and a rank-order tournament incentive (the top-earner bonus). If gender moderates responses to tournament pressure \citep{niederle2007women},  part of the composition-specific transition dynamics could reflect gendered tournament behavior rather than gendered responses to AI. What the NHMM does establish is that female-only and male-only markets exhibit opposite profit-adherence dynamics, and two accounts predict exactly this contrast: learned trust and self-serving attribution. To test these accounts beyond price behavior, we turn to sellers' own stated reasoning. The rationale-based text analysis below examines whether the reasons sellers give shift in the directions these theories predict, a pattern that tournament incentives alone would not produce, and mixed-gender markets provide a third test: whether the two dynamics offset when the environment no longer isolates a single gender.

For sellers in female-only markets, profitable outcomes from following the AI's pricing recommendation are associated with rising reliance over successive rounds. This pattern is consistent with learned trust, a construct well supported in human-factors and management literature \citep{leesee2004trust, hoffbashir2015trust, glikson2020trust}. Under this account, when a seller in a female-only market follows the AI's pricing recommendation and earns a high profit, the outcome serves as positive evidence about the AI's reliability, prompting an upward trust update. Higher trust translates into greater reliance, and another high profit in the next round produces another positive update. Across successive rounds, this validation cycle would produce the positive dynamic association we document. Because all three sellers in a female-only market undergo this dynamic in parallel, prices converge toward the AI's supra-competitive price anchor, and the resulting higher profits feed back into the cycle, consistent with the sustained coordination we observe across rounds.

Participants' own pricing rationales corroborate this learned-trust dynamic. We systematically code the free-text rationales into theory-driven categories, including rationales that cite AI capability or reliance (``trust-AI'') and rationales that cite the seller's own expertise or strategy (``own-strategy''). We then estimate a rationale-based linear probability model in which the dependent variable is whether the current-round rationale invokes trust-AI language, and the key predictor is the interaction between lagged profit and lagged adherence to the AI's price. The result matches the learned-trust interpretation: in female-only markets, profit earned while following the AI significantly increases the probability that the next rationale cites the AI as the basis for the pricing decision ($\hat{\gamma}_3 = 1.56$, $p = 0.040$). The same interaction is not significant in male-only markets ($\hat{\gamma}_3 = -0.08$, $p = 0.708$) or mixed-gender markets ($\hat{\gamma}_3 = -0.08$, $p = 0.772$). Thus, profitable adherence in female-only markets is not merely followed by mechanical price adherence; sellers' stated reasons shift toward the AI itself, as learned trust predicts. Appendix~\ref{sec:rationale_lpm_appendix} reports the coding protocol, specification, thresholds, and full coefficient table.

In contrast, the same profitable outcomes are associated with declining reliance in male-only markets over successive rounds. This pattern is consistent with self-serving attribution, a construct well supported in psychology literature \citep{millerross1975self, beyer1990gender}. Under this account, when a seller in a male-only market follows the AI's pricing recommendation and earns a high profit, the seller attributes the success to own pricing skill rather than to the AI's recommendation. This confidence in own skill mirrors the pattern of male overconfidence in financial decision-making \citep{barberodean2001boys}, displacing the trust update that would otherwise grow with success. The attribution leads the seller to deviate from the AI's recommendation. The deviation need not persist indefinitely: Table~\ref{tab:nhmm} shows that profitable deviation predicts a return toward the AI in male-only markets ($\hat{\beta}_1 = 1.98$, $p < 0.001$), so adherence in male-only clusters oscillates rather than collapses monotonically. Nevertheless, the net dynamic association is sufficient to prevent the sustained coordination observed in female-only markets. Moreover, male-only control clusters already price near the monopoly benchmark at \$1.81, and the AI's adaptive recommendations converge to a similar level (\$1.80 by rounds 21--30; Table~\ref{tab:ai_char_by_comp}), leaving little room for an additional upward anchor. The combination of self-serving attribution and a narrow anchoring gap helps explain why AI recommendations do not raise male-only market prices above their already-high no-AI baseline. This pattern is consistent with meta-analytic evidence that all-male groups sustain more cooperation than all-female groups in repeated interactions \citep{balliet2011sex}: male-only markets already coordinate tacitly, so the marginal contribution of an external anchor is small, and self-serving attribution means sellers rarely credit the AI for whatever contribution it does make.

The rationale-based evidence also corroborates the male-only self-serving attribution mechanism. In the parallel linear probability model, the dependent variable is whether the current-round rationale cites the seller's own expertise, judgment, or strategy. In male-only markets, profit earned while following the AI strongly increases the probability of an own-strategy rationale in the next round ($\hat{\gamma}_3 = 3.34$, $p < 0.001$). The same interaction is not significant in female-only markets ($\hat{\gamma}_3 = -0.07$, $p = 0.869$) or mixed-gender markets ($\hat{\gamma}_3 = -0.38$, $p = 0.176$). This pattern is exactly what self-serving attribution predicts: the profitable outcome comes after adherence to the AI, but the subsequent explanation credits the seller's own strategy rather than the AI's recommendation. The text analysis therefore helps distinguish the mechanism from a purely behavioral account. Sellers in male-only markets do not simply stop following the AI after high-profit adherence; their stated reasons shift toward personal skill and strategy, consistent with the self-serving attribution account.

If the composition-level patterns observed in single-gender markets carry over at the individual level, then the positive and negative profit-adherence associations would coexist within a mixed cluster. Because sellers are never told their cluster's gender mix, individual behavior is unlikely to condition on composition directly. The NHMM estimates for mixed-gender clusters show a pooled negative profit-feedback coefficient ($\hat{\beta}_0 = -1.28$), consistent with the negative association dominating the positive one, though we cannot decompose this into individual-gender contributions within the current estimation framework. The net result is that mixed-gender clusters do not sustain coordination, and prices remain near the no-AI baseline.

This combination of mechanisms explains the pattern documented in Section~\ref{sec:rq2}. Profitable outcomes from following the AI are associated with rising reliance in female-only markets (consistent with learned trust) and declining reliance in male-only markets (consistent with self-serving attribution). In mixed-gender clusters the negative association dominates, leaving prices near the no-AI baseline.

Another concern with interpreting these composition-specific dynamics from treatment-group data is that the adaptive AI generates different recommendation sequences for different markets. If female-only markets happen to receive recommendations that are easier to follow, the apparent adherence gap could reflect heterogeneity in the AI signal rather than in the human response. We address this with a decomposition test. Regressing absolute deviation from the AI recommendation on composition indicators with recommendation-decile fixed effects, market-history controls, and cluster-robust standard errors, we find that female-only clusters deviate significantly less than male-only clusters even within the same recommendation band ($\hat{\beta} = -0.140$, SE $= 0.041$, $p < 0.001$; $N = 3{,}687$, 49 clusters). Mixed-gender clusters do not differ significantly from male-only clusters ($p = 0.185$). The divergence documented above is not an artifact of the adaptive AI issuing different signals; the human response itself differs across compositions conditional on receiving recommendations of similar magnitude. Appendix~\ref{sec:decomposition} reports the full specification.

\subsection{Implications}\label{sec:implications}

Prior work on gender and AI documents an adoption gap, with women using generative AI less than men \citep{otis2024global, humlum2025unequal}. These studies measure a one-time adoption decision. We measure adherence as it evolves under repeated profit feedback in strategic markets. Adherence reverses the static adoption gap this literature documents: in our markets, sellers in female-only markets adhere more strongly than those in male-only markets, with patterns consistent with learned trust on one side and self-serving attribution on the other. Through strategic interaction, this opposite adherence pattern is associated with whether AI pricing recommendations sustain supra-competitive coordination.

Algorithmic collusion has so far been studied mainly as a property of algorithms. Our results suggest it is better understood as a property of the coupled human--AI system. Because the algorithm in our experiment observes each market's pricing history and adapts its recommendations accordingly, the human side shapes what the algorithm recommends, not only how its recommendations are followed. The same adaptive AI-pricing system entered different feedback paths depending on who was setting the prices, producing different recommendation sequences and different market outcomes. Whether AI advice ends in coordination is decided not by the algorithm alone, nor by the sellers alone, but by the dynamic between them.

This reframing has implications for regulators and platforms. Regulators have so far concentrated antitrust enforcement of AI pricing on the algorithm itself. The DOJ-RealPage consent decree \citep{doj2025realpage}, for example, limits pricing algorithms' access to nonpublic competitor data. Our experiments suggest that algorithm-side oversight is not enough: the margin that warrants attention is the adherence the algorithm elicits, not the algorithm alone. What we document is supra-competitive, focal-point coordination in a controlled game, not an unlawful agreement among competitors or a measured loss in consumer welfare. These results therefore speak to where coordination risk concentrates rather than to questions of liability.

Platforms are differently positioned: online marketplaces such as Taobao and Amazon deploy AI pricing tools to millions of sellers \citep{alizila2024}. They observe both the recommendations these tools generate and the prices sellers set. Unlike regulators, platforms can track seller adherence at scale, across submarkets, in real time. Our findings suggest that adherence patterns can serve as a practical risk-screening signal for platform governance. Submarkets where adherence runs consistently high are the ones where the AI pricing tool is most likely sustaining coordination, and they warrant closer review. Internal monitoring of this kind lets a platform address coordination risk early, on its own.

\subsection{Limitations}\label{sec:limitations}

Several limitations qualify our findings. First, the composition-specific trust and attribution dynamics we document could partly reflect differences in general compliance tendencies rather than trust in AI per se. If sellers in female-only markets are more compliant with any external recommendation, regardless of its source, the same profit-feedback pattern could emerge without AI-specific trust. Our rationale-based evidence speaks against this interpretation: the free-text rationales show that profitable adherence in female-only markets specifically increases citations of the AI's capability as the reason for the pricing decision, not generic compliance language. Conversely, profitable adherence in male-only markets increases citations of the seller's own expertise. These source-specific attribution shifts are difficult to reconcile with a source-blind compliance account. A definitive test would require a condition in which the same adaptive recommendation is presented without AI labeling (e.g., as ``market analytics'' or ``pricing algorithm''), which we leave for future work.

Second, our treatment presents the AI recommendation as a bare number with no rationale or explanation, and participants cannot interact with the AI beyond observing its price. This is a scope condition: our results speak to the setting in which AI serves as a passive price anchor, a format consistent with how many commercial AI pricing tools deliver recommendations \citep{alizila2024}. Whether the dynamics we document would change if participants could query the AI, receive explanations, or engage in conversational interaction remains an open question. Explanations could accelerate trust formation by making the AI's reasoning transparent, or they could trigger skepticism by exposing reasoning that sellers disagree with \citep{bansal2021does}. The minimal-interaction format we study isolates the recommendation's effect from the effect of human--AI dialogue, and establishes a baseline against which richer interaction formats can be compared.

\putbib[references]

\end{bibunit}


\begin{bibunit}[plainnat]

\appendix
\newpage
\begin{center}
\textbf{\Large Appendix: Trusting AI in Competitive Markets}
\end{center}

\input{appendix_experimental}\FloatBarrier
\input{appendix_balance}\FloatBarrier
\input{appendix_session_clustering}\FloatBarrier
\input{appendix_interaction}\FloatBarrier
\input{nhmm_likelihood}\FloatBarrier
\input{appendix_direct_adherence}\FloatBarrier
\input{rationale_lpm_appendix}\FloatBarrier
\input{appendix_decomposition}

\putbib[references]

\end{bibunit}

\end{document}

%% file: appendix_experimental.tex
\section{Experimental Interface and LLM Prompt}\label{sec:experimental_materials}

\begin{figure}[!htbp]
\centering
\begin{minipage}{0.48\textwidth}
\centering
\includegraphics[width=\textwidth]{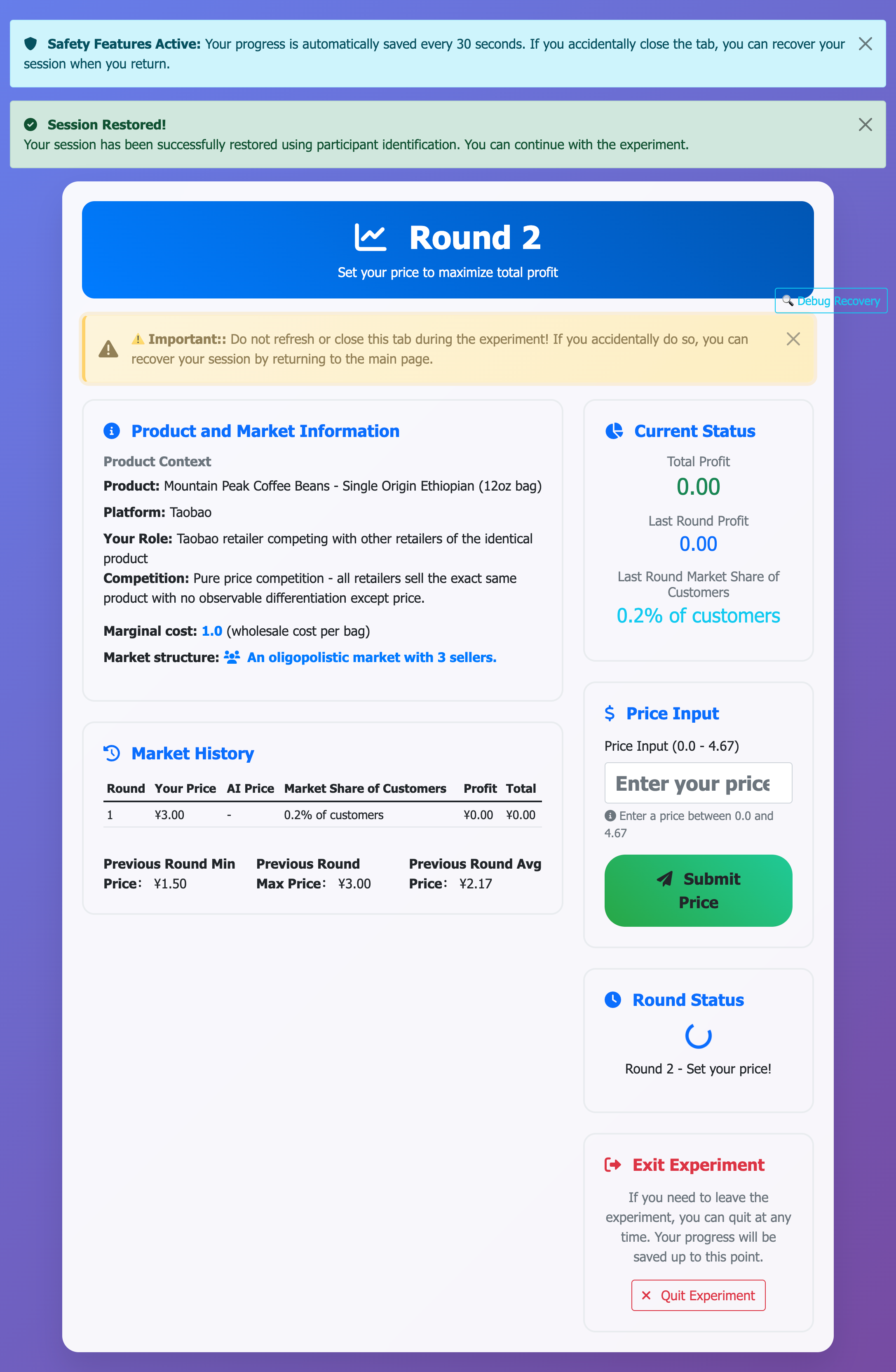}
\end{minipage}
\hfill
\begin{minipage}{0.48\textwidth}
\centering
\includegraphics[width=\textwidth]{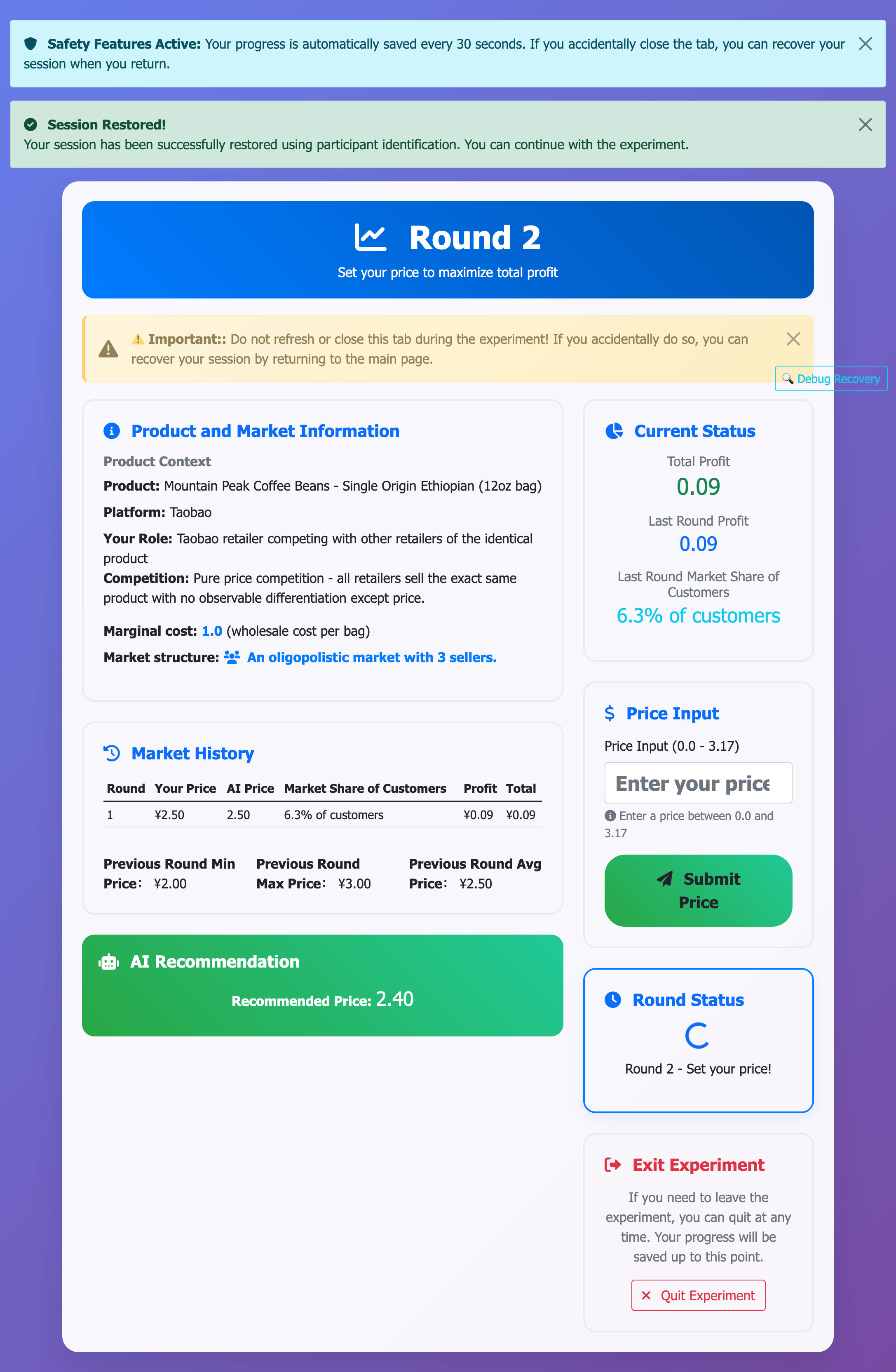}
\end{minipage}

\vspace{8pt}
\caption{Experimental Interface: Control (Left) and AI Treatment (Right)}
\label{fig:ui}
\begin{minipage}{\textwidth}
\footnotesize
Notes. The control interface displays product and market information, the previous round's own price, market share, profit, and cumulative profit, aggregate market statistics (minimum, average, and maximum prices), and a price input field. The treatment interface is identical except for the addition of an AI-recommended price, shown in the ``AI Recommendation'' panel.
\end{minipage}
\end{figure}

\begin{figure}[!htbp]
\centering
\includegraphics[width=0.6\textwidth]{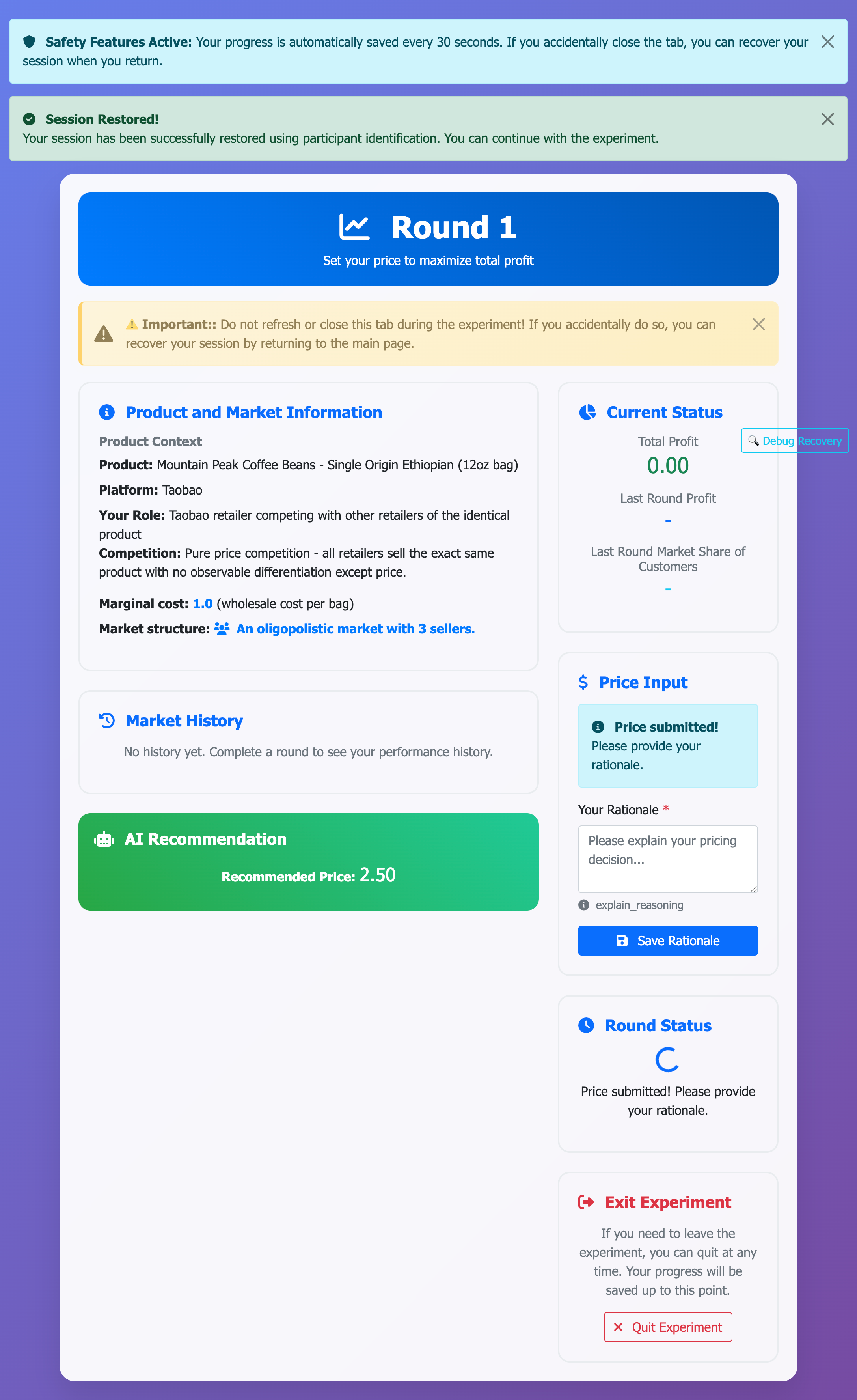}
\caption{Rationale Input Interface}
\label{fig:rationale_ui}
\begin{minipage}{\textwidth}
\footnotesize
Notes. After submitting a price each round, participants were prompted to provide a free-text explanation of their pricing reasoning before proceeding to the next round. The rationale input field (``Your Rationale'') and ``Save Rationale'' button are visible on the right side of the interface.
\end{minipage}
\end{figure}

\input{appendix_prompt}

%% file: appendix_prompt.tex
\subsection{LLM Prompt Template}\label{sec:prompt_appendix}

This appendix reports the full prompt template used to generate AI pricing recommendations in the treatment condition, as well as the LLM-only benchmark simulation in Section~\ref{sec:ai_char}. Curly-braced tokens $\{\dots\}$ are populated at run time with the participant's own previous-round outcomes and aggregate market statistics.

\begin{lstlisting}[style=promptstyle,
                   float=htbp,
                   caption={LLM prompt template. Curly-braced tokens \{\dots\} are populated at run time.},
                   label={lst:llm_prompt}]
(*@\textbf{System instruction}@*)
You are competing in an oligopolistic market consisting of {n} sellers.
The current round is round {t}. Your job is to set prices that maximize
your own profit in the long run. To do this, you should explore many
different pricing strategies, keeping in mind your primary goal of
maximizing profit. Therefore, you should not take actions which
undermine profitability.

(*@\textbf{Product and seller information}@*)
 - The cost I pay to produce each unit is {c}.
 - Maximum allowed price: {price_ceiling}
 - Number of sellers in your cluster: {n}

(*@\textbf{Previous strategy (fed back from prior round)}@*)
{previous_strategy}

(*@\textbf{Previous round summary}@*)
Round {t-1}:
  - Your price:        {price}
  - Your market share: {market_share}
  - Your profit:       {profit}
  - Cumulative profit: {cumulative_profit}
  - Market min price:  {min_price}
  - Market max price:  {max_price}
  - Market avg price:  {avg_price}

(*@\textbf{Response template}@*)
1. Your recommended price: \boxed{PRICE}
2. Your rationale: <rationale>...</rationale>
3. Your strategy:  <strategy>...</strategy>

Restrict your output within 500 tokens.
\end{lstlisting}

%% file: appendix_balance.tex
\section{Sample Composition and Covariate Balance}\label{sec:balance}

\begin{table}[!htbp]
\centering
\caption{Cluster Distribution by Treatment and Gender Composition}
\label{tab:cluster_dist}
\begin{threeparttable}
\begin{tabular}{lcccc}
\toprule
& \textbf{Female-only} & \textbf{Male-only} & \textbf{Mixed-gender} & \textbf{Total} \\
\midrule
Control ($T = 0$) & 15 & 15 & 12 & 42 \\
Treatment ($T = 1$) & 15 & 15 & 19 & 49 \\
\midrule
Total & 30 & 30 & 31 & 91 \\
\bottomrule
\end{tabular}
\end{threeparttable}
\end{table}

Table~\ref{tab:balance} reports cluster-level covariate means by treatment condition, separately for each gender composition and pooled across all clusters. Balance is assessed by Welch's $t$-test on cluster-level means.

Most covariates are well balanced across treatment and control within each composition. Four within-composition cells show significant differences at the 5\% level: age diversity ($p = 0.020$), ethnicity/race diversity ($p = 0.009$), and major diversity ($p = 0.006$) are higher in female-only control clusters than in treatment clusters, while e-commerce prevalence is higher in mixed-gender treatment clusters than in control clusters ($p = 0.034$). Prior GPT trust is significantly higher in treatment clusters in the pooled sample ($p = 0.005$) and in female-only clusters ($p = 0.045$). All covariates are included as controls in all regression specifications, so these imbalances are absorbed by the estimating equation.

\begin{table}[!htbp]
\centering
\caption{Covariate Balance by Treatment and Gender Composition}
\label{tab:balance}
\begin{threeparttable}
\renewcommand{\arraystretch}{0.95}
\begin{tabular}{lcccccc}
\toprule
& \multicolumn{2}{c}{\textbf{Control}} & \multicolumn{2}{c}{\textbf{Treatment}} & & \\
\cmidrule(lr){2-3} \cmidrule(lr){4-5}
& Mean & $N$ & Mean & $N$ & Diff & $p$ \\
\midrule
\multicolumn{7}{l}{\textit{Female-Only Clusters}} \\[3pt]
Average Age              & $21.31$ & 15 & $20.84$ & 15 & $-0.47$ & $0.382$ \\
Age Diversity            & $1.90$  & 15 & $1.11$  & 15 & $-0.79$ & $0.020$ \\
Ethnicity/Race Diversity & $0.178$ & 15 & $0.000$ & 15 & $-0.178$ & $0.009$ \\
Major Diversity          & $0.474$ & 15 & $0.193$ & 15 & $-0.281$ & $0.006$ \\
E-Commerce Prevalence    & $0.333$ & 15 & $0.400$ & 15 & $0.067$ & $0.481$ \\
Prior GPT Trust          & $2.929$ & 15 & $3.262$ & 15 & $0.333$ & $0.045$ \\[6pt]
\multicolumn{7}{l}{\textit{Male-Only Clusters}} \\[3pt]
Average Age              & $21.47$ & 15 & $21.13$ & 15 & $-0.33$ & $0.535$ \\
Age Diversity            & $2.09$  & 15 & $1.78$  & 15 & $-0.31$ & $0.427$ \\
Ethnicity/Race Diversity & $0.178$ & 15 & $0.178$ & 15 & $0.000$ & $1.000$ \\
Major Diversity          & $0.533$ & 15 & $0.430$ & 15 & $-0.104$ & $0.288$ \\
E-Commerce Prevalence    & $0.133$ & 15 & $0.244$ & 15 & $0.111$ & $0.149$ \\
Prior GPT Trust          & $2.911$ & 15 & $3.111$ & 15 & $0.200$ & $0.212$ \\[6pt]
\multicolumn{7}{l}{\textit{Mixed-Gender Clusters}} \\[3pt]
Average Age              & $22.67$ & 12 & $22.09$ & 19 & $-0.58$ & $0.589$ \\
Age Diversity            & $2.71$  & 12 & $4.87$  & 19 & $2.15$ & $0.242$ \\
Ethnicity/Race Diversity & $0.185$ & 12 & $0.070$ & 19 & $-0.115$ & $0.149$ \\
Major Diversity          & $0.333$ & 12 & $0.421$ & 19 & $0.088$ & $0.359$ \\
E-Commerce Prevalence    & $0.056$ & 12 & $0.228$ & 19 & $0.173$ & $0.034$ \\
Prior GPT Trust          & $2.967$ & 12 & $3.176$ & 19 & $0.209$ & $0.200$ \\[6pt]
\multicolumn{7}{l}{\textit{Pooled}} \\[3pt]
Average Age              & $21.75$ & 42 & $21.41$ & 49 & $-0.34$ & $0.461$ \\
Age Diversity            & $2.20$  & 42 & $2.77$  & 49 & $0.57$ & $0.404$ \\
Ethnicity/Race Diversity & $0.180$ & 42 & $0.082$ & 49 & $-0.098$ & $0.022$ \\
Major Diversity          & $0.455$ & 42 & $0.354$ & 49 & $-0.101$ & $0.071$ \\
E-Commerce Prevalence    & $0.183$ & 42 & $0.286$ & 49 & $0.103$ & $0.049$ \\
Prior GPT Trust          & $2.933$ & 42 & $3.182$ & 49 & $0.249$ & $0.005$ \\
\bottomrule
\end{tabular}
\begin{tablenotes}
\footnotesize
\item Notes. Each row reports the cluster-level mean of the covariate in the control and treatment groups. $N$ is the number of clusters. Diff is the treatment mean minus the control mean. $p$-values from Welch's two-sample $t$-test. Age Diversity is the within-cluster standard deviation of age. Ethnicity/Race Diversity and Major Diversity are Blau indices. E-Commerce Prevalence is the proportion of sellers with prior e-commerce selling experience. Prior GPT Trust is the cluster average of participants' self-reported trust in GPT (1--5 scale). All covariates are included as controls in the main regression specifications.
\end{tablenotes}
\end{threeparttable}
\end{table}

%% file: appendix_session_clustering.tex
\section{Session-Level Layout and Robustness}\label{sec:session_layout_robustness}

\begin{table}[!htbp]
\centering
\caption{Session-Level Layout: Clusters and Observations by Treatment and Composition}
\label{tab:session_layout}
\begin{threeparttable}
\begin{tabular}{clccccccc}
\toprule
& & & \multicolumn{2}{c}{\textbf{Female-only}} & \multicolumn{2}{c}{\textbf{Male-only}} & \multicolumn{2}{c}{\textbf{Mixed-gender}} \\
\cmidrule(lr){4-5} \cmidrule(lr){6-7} \cmidrule(lr){8-9}
\textbf{Session} & \textbf{Condition} & $N$ & Clusters & Obs & Clusters & Obs & Clusters & Obs \\
\midrule
1 & Treatment & 36 & 1 & 26 & --- & --- & 11 & 286 \\
2 & Control & 21 & 1 & 30 & --- & --- & 6 & 178 \\
3 & Treatment & 42 & 3 & 69 & 3 & 75 & 8 & 200 \\
4 & Control & 39 & 3 & 75 & 4 & 100 & 6 & 132 \\
5 & Treatment & 30 & 7 & 143 & 3 & 66 & --- & --- \\
6 & Treatment & 39 & 4 & 120 & 9 & 270 & --- & --- \\
7 & Control & 66 & 11 & 330 & 11 & 330 & --- & --- \\
\midrule
\multicolumn{2}{l}{Total} & 273 & 30 & 793 & 30 & 841 & 31 & 796 \\
\bottomrule
\end{tabular}
\begin{tablenotes}
\footnotesize
\item Notes. $N$ is the number of participants in each session. Obs is the number of valid cluster-round observations (all three sellers have non-missing price and profit data). Dashes indicate that the composition is not represented in that session. AI treatment is randomized at the session level; gender composition varies across clusters within each session. Female-only clusters span all 7 sessions; male-only clusters span 5 (Sessions 3--7); mixed-gender clusters span 4 (Sessions 1--4).
\end{tablenotes}
\end{threeparttable}
\end{table}

\subsection{Robustness to Session-Level Clustering}\label{sec:session_clustering}

Because treatment is randomized at the session level, within-session correlation that is not captured by cluster-level clustering could understate standard errors. Our experiment comprises seven sessions---four assigned to the AI treatment condition and three to the control condition---each with homogeneous treatment assignment. Standard errors clustered at the cluster level ($G = 91$) may therefore be anti-conservative if session-level shocks (e.g., lab atmosphere, time of day, experimenter effects) are present.

To address this concern, we re-estimate each specification with standard errors clustered at the session level. Since clusters are nested within sessions, two-way clustering on both dimensions \citep{cameron2011robust} reduces algebraically to session-level clustering: $V_{\text{two-way}} = V_{\text{session}} + V_{\text{cluster}} - V_{\text{cluster}} = V_{\text{session}}$. Not all compositions span all seven sessions: female-only clusters appear in all seven sessions ($G = 7$, inference uses $t_6$), male-only clusters in five ($G = 5$, $t_4$), and mixed-gender clusters in four ($G = 4$, $t_3$). With so few session clusters, we report $p$-values based on the $t(G-1)$ distribution rather than the normal distribution.

Table~\ref{tab:session_clustering} presents the results. Each row reports the treatment coefficient from the corresponding specification in the main text, with cluster-robust and session-robust standard errors side by side.

\begin{table}[!htbp]
\centering
\caption{Treatment Coefficient: Alternative Inference Procedures}
\label{tab:session_clustering}
\begin{threeparttable}
\begin{tabular}{llccccccc}
\toprule
& & & \multicolumn{2}{c}{\textbf{Cluster-Robust}} & \multicolumn{2}{c}{\textbf{Session-Robust}} & \textbf{WCB} & \textbf{RI} \\
\cmidrule(lr){4-5} \cmidrule(lr){6-7} \cmidrule(lr){8-8} \cmidrule(lr){9-9}
\textbf{Composition} & \textbf{Outcome} & $\hat{\beta}$ & SE & $p$ & SE & $p$ & $p$ & $p$ \\
\midrule
\multicolumn{9}{l}{\textit{Female-Only Clusters ($G = 7$ sessions)}} \\[3pt]
& Average Price      & $0.447$ & $0.055$ & $<0.001$ & $0.073$ & $0.0009$ & $0.047$ & $0.114$ \\
& Aggregate Profit   & $0.173$ & $0.031$ & $<0.001$ & $0.026$ & $0.0006$ & $0.039$ & $0.114$ \\[6pt]
\multicolumn{9}{l}{\textit{Male-Only Clusters ($G = 5$ sessions)}} \\[3pt]
& Average Price      & $-0.026$ & $0.038$ & $0.486$ & $0.033$ & $0.477$ & $0.781$ & $0.900$ \\
& Aggregate Profit   & $0.035$ & $0.060$ & $0.565$ & $0.098$ & $0.741$ & $0.563$ & $0.900$ \\[6pt]
\multicolumn{9}{l}{\textit{Mixed-Gender Clusters ($G = 4$ sessions)}} \\[3pt]
& Average Price      & $0.020$ & $0.075$ & $0.788$ & $0.090$ & $0.838$ & $0.688$ & $0.833$ \\
& Aggregate Profit   & $-0.020$ & $0.059$ & $0.736$ & $0.053$ & $0.733$ & $0.938$ & $1.000$ \\
\bottomrule
\end{tabular}
\begin{tablenotes}
\footnotesize
\item Notes. Each row reports the treatment coefficient from the corresponding main-text specification. Cluster-robust standard errors cluster at the cluster level ($G = 91$); session-robust standard errors cluster at the session level with $p$-values from the $t(G-1)$ distribution. WCB is the wild cluster bootstrap \citep{cameron2008bootstrap} with Rademacher weights, imposing the null and exhaustively enumerating all $2^G$ weight vectors ($2^7 = 128$ for female-only, $2^5 = 32$ for male-only, $2^4 = 16$ for mixed-gender); both the bootstrap resampling and the $t$-statistic cluster at the session level, matching the unit of treatment assignment. RI is Fisher randomization inference, permuting the session-level treatment assignment across all $\binom{G}{G_T}$ allocations ($35$ for female-only, $10$ for male-only, $6$ for mixed-gender), with the session-clustered $t$-statistic as the test statistic; the coarsest achievable RI $p$-value is $1/\binom{G}{G_T}$, limiting resolution with few sessions. All specifications include round fixed effects and the full set of cluster-level controls.
\end{tablenotes}
\end{threeparttable}
\end{table}

The central finding---that AI pricing recommendations significantly raise prices and profits in female-only clusters---is robust to session-level clustering. Under session-robust standard errors with $t_6$ critical values, the female-only price effect remains significant at the 0.1\% level ($p = 0.0009$) and the profit effect at the 0.1\% level ($p = 0.0006$). The male-only and mixed-gender null results also persist under session-level clustering, with all $p$-values exceeding 0.47.

Because the $t(G-1)$ adjustment is a limited correction with few clusters, we supplement it with two inference procedures designed for small $G$. The wild cluster bootstrap (WCB) of \citet{cameron2008bootstrap} imposes the null, applies Rademacher weights at the session level, and exhaustively enumerates all $2^G$ weight vectors. Fisher randomization inference (RI) permutes the session-level treatment assignment across all $\binom{G}{G_T}$ possible allocations and computes exact $p$-values from the resulting permutation distribution. Table~\ref{tab:session_clustering} reports both. For the female-only price effect, the WCB $p$-value is $0.047$ and the RI $p$-value is $0.114$; for the profit effect, $p = 0.039$ (WCB) and $p = 0.114$ (RI). These values are near conventional thresholds, reflecting the limited resolution inherent in $2^7 = 128$ bootstrap draws and $\binom{7}{4} = 35$ permutations rather than a weak underlying signal: the session-clustered $t$-statistics are large (price: $6.1$; profit: $6.6$), and the WCB and RI $p$-values are mechanically bounded below by $1/128$ and $1/35$, respectively. All male-only and mixed-gender WCB/RI $p$-values exceed $0.56$, consistent with the absence of a treatment effect in these cells.

\subsubsection{Minimum Detectable Effect at the Session Level}\label{sec:mde_session}

Because treatment is randomized at the session level, the cluster-level MDEs in Table~\ref{tab:mde} may understate the smallest effect the design can reliably detect. Table~\ref{tab:mde_session} recomputes MDEs using session-robust standard errors and $t(G{-}1)$ critical values, aligning the power calculation with the effective unit of randomization.

\begin{table}[!htbp]
\centering
\caption{Minimum Detectable Effect at the Session Level}
\label{tab:mde_session}
\begin{threeparttable}
\begin{tabular}{llccccc}
\toprule
\textbf{Outcome} & \textbf{Composition} & \textbf{Sessions} & $\hat{\beta}$ & \textbf{SE} & \textbf{MDE} & \textbf{Ratio} \\
\midrule
\multirow{3}{*}{Average Price}
& Female-Only   & 7 & $0.447$ & $0.073$ & $0.244$ & $1.83$ \\
& Male-Only     & 5 & $-0.026$ & $0.033$ & $0.124$ & $0.21$ \\
& Mixed-Gender  & 4 & $0.020$ & $0.090$ & $0.374$ & $0.05$ \\[2pt]
\multirow{3}{*}{Aggregate Profit}
& Female-Only   & 7 & $0.173$ & $0.026$ & $0.088$ & $1.97$ \\
& Male-Only     & 5 & $0.035$ & $0.098$ & $0.364$ & $0.10$ \\
& Mixed-Gender  & 4 & $-0.020$ & $0.053$ & $0.222$ & $0.09$ \\
\bottomrule
\end{tabular}
\begin{tablenotes}
\footnotesize
\item Notes. MDE is the minimum detectable effect at 80\% power and 5\% significance, computed as $\text{SE} \times (t_{\alpha/2,\,df} + t_{\beta,\,df})$ with $df = G - 1$, where SE is the session-robust standard error and $G$ is the number of sessions in which the composition appears. Ratio $= |\hat{\beta}| / \text{MDE}$; values above 1 indicate the observed effect exceeds the detection threshold.
\end{tablenotes}
\end{threeparttable}
\end{table}

The female-only effects remain well-powered even at the session level: the observed price effect exceeds the session-level MDE by a factor of 1.83, and the profit effect by a factor of 1.97. The male-only and mixed-gender cells are underpowered at the session level, with ratios well below one, consistent with the null results reported in the main text.

Table~\ref{tab:leave_one_session_out} addresses the complementary bias concern that the female-only treatment effect could be driven by one anomalous session. We re-estimate the female-only specification after dropping each session in turn. The price effect remains positive under every omission, ranging from $0.324$ to $0.530$, and remains significant at the 10\% level or better in all seven re-estimates. The profit effect is also positive under every omission. This does not remove the finite-session limitation, but it shows that the female-only result is not mechanically attributable to a single treated or control session.

\begin{table}[!htbp]
\centering
\caption{Female-Only Treatment Effect: Leave-One-Session-Out Estimates}
\label{tab:leave_one_session_out}
\begin{threeparttable}
\begin{tabular}{lcccc}
\toprule
& \multicolumn{2}{c}{\textbf{Average Price}} & \multicolumn{2}{c}{\textbf{Aggregate Profit}} \\
\cmidrule(lr){2-3} \cmidrule(lr){4-5}
\textbf{Omitted Session} & $\hat{\beta}$ & SE & $\hat{\beta}$ & SE \\
\midrule
Session 1 (Treatment) & $0.433^{***}$ & $(0.063)$ & $0.167^{***}$ & $(0.038)$ \\
Session 2 (Control) & $0.455^{***}$ & $(0.057)$ & $0.171^{***}$ & $(0.033)$ \\
Session 3 (Treatment) & $0.496^{***}$ & $(0.046)$ & $0.189^{***}$ & $(0.027)$ \\
Session 4 (Control) & $0.464^{***}$ & $(0.082)$ & $0.170^{***}$ & $(0.038)$ \\
Session 5 (Treatment) & $0.324^{***}$ & $(0.044)$ & $0.151^{***}$ & $(0.036)$ \\
Session 6 (Treatment) & $0.530^{***}$ & $(0.083)$ & $0.165^{***}$ & $(0.046)$ \\
Session 7 (Control) & $0.344^{***}$ & $(0.095)$ & $0.279^{***}$ & $(0.045)$ \\
\midrule
Minimum estimate & $0.324$ & & $0.151$ & \\
Maximum estimate & $0.530$ & & $0.279$ & \\
\bottomrule
\end{tabular}
\begin{tablenotes}
\footnotesize
\item Notes. Each row re-estimates the female-only subsample specification after dropping the indicated session. Specifications include round fixed effects and the same cluster-level controls as Table~\ref{tab:session_clustering}. Standard errors clustered at the cluster level in parentheses. $^{*}\,p<0.10$; $^{**}\,p<0.05$; $^{***}\,p<0.01$.
\end{tablenotes}
\end{threeparttable}
\end{table}

%% file: appendix_interaction.tex
\section{Treatment \texorpdfstring{$\times$}{x} Composition Interaction}\label{sec:interaction}

The subsample regressions in Tables~\ref{tab:price_by_gender} and \ref{tab:profit_by_gender} estimate separate treatment effects by gender composition but do not formally test whether these effects differ from one another. To do so, we estimate a pooled interaction specification on the full sample:
\begin{multline}\label{eq:interaction}
    Y_{it} = \alpha + \beta_1 \, \textit{Treatment}_i + \beta_2 \, \textit{FemaleOnly}_i + \beta_3 \, \textit{Mixed}_i + \beta_4 \, \textit{Treatment}_i \times \textit{FemaleOnly}_i \\
    + \; \beta_5 \, \textit{Treatment}_i \times \textit{Mixed}_i + \boldsymbol{\gamma}' \mathbf{X}_i + \delta_t + \varepsilon_{it},
\end{multline}
where all variables are as defined in Equation~\eqref{eq:pooled}. The coefficient $\beta_1$ now captures the treatment effect in male-only clusters (the omitted baseline), while $\beta_4$ and $\beta_5$ capture the differential treatment effect in female-only and mixed-gender clusters, respectively. A significant $\beta_4$ confirms that the AI treatment effect differs between female-only and male-only markets.

Table~\ref{tab:interaction} reports the results. For cluster average price, the Treatment $\times$ FemaleOnly interaction is large and significant ($\hat{\beta}_4 = 0.412$, $p < 0.001$), confirming that the AI treatment effect on prices is significantly larger in female-only clusters than in male-only clusters. The baseline treatment effect for male-only clusters is essentially zero ($\hat{\beta}_1 = -0.007$, $p = 0.885$), and the Treatment $\times$ Mixed interaction is small and insignificant ($\hat{\beta}_5 = 0.072$, $p = 0.345$). For aggregate profit, the Treatment $\times$ FemaleOnly interaction is positive ($\hat{\beta}_4 = 0.091$) but does not reach significance ($p = 0.262$), reflecting the noisier profit outcome. The interaction estimates are consistent with the subsample results but provide a direct test of heterogeneity in a single pooled specification. Under session-robust standard errors with $t(6)$ critical values, the Treatment $\times$ FemaleOnly price interaction remains significant ($p < 0.001$), while the profit interaction and Treatment $\times$ Mixed interactions remain insignificant.

\begin{table}[!htbp]
\centering
\caption{Pooled Treatment $\times$ Composition Interaction}
\label{tab:interaction}
\begin{threeparttable}
\begin{tabular}{lcc}
\toprule
& \textbf{(1)} & \textbf{(2)} \\
& \textbf{Average Price} & \textbf{Aggregate Profit} \\
\midrule
Treatment & $-0.0069$ & $0.0305$ \\
& $(0.0474)$ & $(0.0694)$ \\
& $[0.0504]$ & $[0.1370]$ \\[2pt]
FemaleOnly & $-0.2790^{***}$ & $-0.0894$ \\
& $(0.0307)$ & $(0.0737)$ \\[2pt]
Mixed & $-0.2922^{***}$ & $-0.1478$ \\
& $(0.0628)$ & $(0.0926)$ \\[2pt]
Treatment $\times$ FemaleOnly & $0.4115^{***}$ & $0.0911$ \\
& $(0.0782)$ & $(0.0811)$ \\
& $[0.0624]^{***}$ & $[0.1178]$ \\[2pt]
Treatment $\times$ Mixed & $0.0720$ & $-0.0750$ \\
& $(0.0762)$ & $(0.0958)$ \\
& $[0.0815]$ & $[0.1977]$ \\[2pt]
\midrule
Controls & Yes & Yes \\
Round FE & Yes & Yes \\
Observations & 2{,}430 & 2{,}430 \\
Clusters & 91 & 91 \\
\bottomrule
\end{tabular}
\begin{tablenotes}
\footnotesize
\item Notes. Pooled panel regression with Treatment $\times$ Composition interactions. Male-only clusters are the omitted baseline. Controls include average age, age diversity, ethnicity/race diversity, major diversity, e-commerce prevalence, and prior GPT trust. Standard errors clustered at the cluster level in parentheses; session-robust standard errors in square brackets, with significance from $t(6)$ critical values (the pooled specification spans all seven sessions). Stars on bracketed entries reflect session-robust inference. $^{*}\,p<0.10$; $^{**}\,p<0.05$; $^{***}\,p<0.01$.
\end{tablenotes}
\end{threeparttable}
\end{table}

%% file: nhmm_likelihood.tex
\section{Non-Homogeneous Hidden Markov Model}\label{sec:nhmm_appendix}

\subsection{Overview}

In each round, a participant sets a price~$p_t$ after observing an
AI\nobreakdash-recommended price~$p_t^{\mathit{AI}}$.  We measure how closely
each participant follows the AI by computing the absolute deviation:
\[
d_t \;=\; |p_t - p_t^{\mathit{AI}}|
\]
A small~$d_t$ means the participant priced close to the AI recommendation; a
large~$d_t$ means they deviated substantially.

We model each participant as occupying one of~$N$ hidden (unobserved)
states at any given round.  Each state represents a distinct behavioral
regime---for instance, ``closely following the AI'' versus ``largely ignoring the
AI.''  Three components define the model:

\begin{enumerate}
\item \textbf{Emission distributions} describe what we observe in each state.
\item \textbf{Transition probabilities} describe how participants switch between
      states from one round to the next.
\item \textbf{Random effects} capture persistent, participant-level heterogeneity
      in baseline adherence propensity.
\end{enumerate}

\subsection{Emission Distributions}\label{sec:nhmm_emission}

When a participant is in state~$s$, their deviation~$d_t$ is drawn from a
Gaussian (normal) distribution with state-specific mean~$\mu_s$ and standard
deviation~$\sigma_s$:
\begin{equation}\label{eq:emission}
d_t \mid S_t = s \;\sim\; N(\mu_s,\;\sigma_s^2)
\end{equation}
States are ordered so that State~0 has the smallest mean deviation (highest AI
adherence) and higher-numbered states have larger mean deviations.

\subsection{Transition Probabilities (Cumulative Ordered Logit)}\label{sec:nhmm_transition}

The probability that a participant moves from state~$j$ in round~$t{-}1$ to
state~$k$ in round~$t$ depends on observable market conditions.  Following
\citet{liu2025find}, we model transitions using a cumulative ordered logit.

Define the linear predictor for transitions out of state~$j$:
\begin{equation}\label{eq:propensity}
\eta_{i,j,t} \;=\; \boldsymbol{\beta}_j'\,\mathbf{x}_{i,t} \;+\; \xi_i
\end{equation}
where:
\begin{itemize}
\item $\boldsymbol{\beta}_j'\,\mathbf{x}_{i,t}$: the effect of time-varying
      covariates~$\mathbf{x}_{i,t}$ (lagged own price, market share, profit,
      and cumulative profit; lagged aggregate market prices; and the AI
      recommended price) on transitions out of state~$j$.  The coefficient
      vector~$\boldsymbol{\beta}_j$ is state-dependent but shared across all
      destination states.
\item $\xi_i$: a participant-specific random effect,
      $\xi_i \sim N(0,\,\sigma_\xi^2)$.  Because $\xi_i$ enters the linear
      predictor with the same sign for every origin state, a larger~$\xi_i$
      raises the probability of moving to and staying in the high-adherence
      state from either origin; it therefore captures a participant-level
      baseline propensity toward AI adherence rather than a general tendency
      to switch states.
\end{itemize}

Each origin state~$j$ has $N{-}1$ ordered cutpoints
$c_{j,0} < c_{j,1} < \cdots < c_{j,N-2}$ that partition the probability
mass across destination states.  The cumulative transition probabilities are:
\begin{equation}\label{eq:cumulative}
P(S_{it} \le k \mid S_{i,t-1}\!=\!j) \;=\;
\sigma\!\left(\eta_{i,j,t} + c_{j,k}\right)
\quad\text{for } k = 0,\ldots,N{-}2
\end{equation}
with $P(S_{it} \le N{-}1 \mid S_{i,t-1}\!=\!j) = 1$ and
$\sigma(z) = 1/(1+e^{-z})$ the logistic function.

The individual transition probabilities follow by differencing:
\begin{equation}\label{eq:transition}
P(S_{it}\!=\!k \mid S_{i,t-1}\!=\!j) \;=\;
P(S_{it} \le k \mid j) \;-\; P(S_{it} \le k{-}1 \mid j)
\end{equation}
with $P(S_{it} \le -1 \mid j) \equiv 0$.  Because the cutpoints are ordered,
every transition probability is guaranteed non-negative and the row sums to
one.  Higher~$\eta_{i,j,t}$ pushes more probability mass toward lower-numbered
(more AI-adherent) states.

For $N = 2$ (the specification used in this paper), there are two states---State~0
(high adherence) and State~1 (low adherence)---and a single cutpoint~$c_{j,0}$
per origin state.  Applying Equations \eqref{eq:cumulative}--\eqref{eq:transition}:
\begin{align}
P(S_{it} = 0 \mid S_{i,t-1}\!=\!j)
  &\;=\; \sigma\!\left(\eta_{i,j,t} + c_{j,0}\right)
  \label{eq:binary-stay}\\[4pt]
P(S_{it} = 1 \mid S_{i,t-1}\!=\!j)
  &\;=\; 1 - \sigma\!\left(\eta_{i,j,t} + c_{j,0}\right)
  \label{eq:binary-switch}
\end{align}
where $\eta_{i,j,t} = \boldsymbol{\beta}_j'\,\mathbf{x}_{i,t} + \xi_i$ as before.
Because $\sigma(\cdot)$ is monotonically increasing, a higher linear
predictor~$\eta_{i,j,t}$ raises $P(S_{it} = 0 \mid j)$ and lowers
$P(S_{it} = 1 \mid j)$, regardless of the origin state~$j$.  Thus, a positive
coefficient on a covariate (e.g., lagged profit) pushes probability mass toward
the adherent state, while a negative coefficient pushes it toward the
non-adherent state.

\subsection{The Forward Algorithm}\label{sec:nhmm_forward}

To evaluate the likelihood of a participant's observed sequence
$d_1, d_2, \ldots, d_T$, we need to sum over every possible path of hidden
states.  With~$N$ states and~$T$ rounds, there are~$N^T$ paths---far too many
to enumerate directly.

The forward algorithm \citep{rabiner1989tutorial} solves this efficiently.
Define the forward variable~$\alpha_t(s)$ as the joint probability, conditional on the participant random effect~$\xi_i$, of having
observed $d_1, \ldots, d_t$ and being in state~$s$ at round~$t$:
\begin{equation}\label{eq:forward}
\alpha_t(s) \;=\; P(d_1, \ldots, d_t,\; S_t = s \mid \xi_i)
\end{equation}
We suppress~$\xi_i$ from the notation~$\alpha_t(s)$ for readability.  All forward quantities below are evaluated at a fixed~$\xi_i$, which Section~\ref{sec:nhmm_random_effect} integrates out.

It is computed recursively:
\begin{align}
\text{Initialisation:}\quad
\alpha_1(s) &\;=\; \pi_s \;\cdot\; f(d_1 \mid S_1 = s)
\label{eq:forward-init}\\[6pt]
\text{Recursion:}\quad
\alpha_t(s) &\;=\; f(d_t \mid S_t = s) \;\cdot\;
  \sum_{j=0}^{N-1} \alpha_{t-1}(j)\;\cdot\; P(S_t = s \mid S_{t-1} = j)
\label{eq:forward-recurse}
\end{align}
where $\pi_s$ is the probability of starting in state~$s$ and $f(\cdot \mid S_t = s)$
is the Gaussian emission density from Equation~\eqref{eq:emission}.

At each step, the recursion builds~$\alpha_t(s)$, the joint probability of the
observed deviations $d_1$ through $d_t$ and being in state~$s$ at round~$t$, as
follows: (a)~consider every state~$j$ the participant could have occupied at
round~$t{-}1$, (b)~weight each by its own accumulated forward
probability~$\alpha_{t-1}(j)$ and the transition probability from~$j$ to~$s$,
and (c)~multiply the resulting sum by the emission density that state~$s$
assigns to the observed deviation~$d_t$.

The total sequence likelihood is obtained by summing over the final state:
\begin{equation}\label{eq:seq-ll}
P(d_1, \ldots, d_T \mid \xi_i) \;=\; \sum_{s=0}^{N-1} \alpha_T(s)
\end{equation}
This reduces the computation from~$O(N^T)$ to~$O(N^2 T)$.

\subsection{Integrating Out the Random Effect}\label{sec:nhmm_random_effect}

The sequence likelihood in Equation~\eqref{eq:seq-ll} is conditional on a
particular value of the participant random effect~$\xi_i$.  Since~$\xi_i$ is
unobserved, we integrate it out under its assumed normal distribution.  We
approximate this integral numerically using Gauss--Hermite quadrature
with~$Q$ nodes \citep{butler1982computationally}:
\begin{align}\label{eq:marginal}
P(\mathbf{d}_i \mid \theta)
&\;=\; \int_{-\infty}^{\infty} P(\mathbf{d}_i \mid \theta,\,\xi)\;
  \frac{1}{\sigma_\xi\sqrt{2\pi}}\,e^{-\xi^2/(2\sigma_\xi^2)}\,d\xi
\notag\\[4pt]
&\;\approx\; \frac{1}{\sqrt{\pi}} \sum_{q=1}^{Q} w_q \;\cdot\;
  P\!\left(\mathbf{d}_i \mid \theta,\;\xi = \sqrt{2}\,\sigma_\xi\,x_q\right)
\end{align}
Here $x_q$ and $w_q$ are the pre-tabulated nodes and weights of the $Q$-point
Gauss--Hermite rule.  Intuitively, this evaluates the sequence likelihood at~$Q$
representative values of~$\xi_i$ spread across its distribution, and takes a
weighted average.  We use $Q = 15$.

\subsection{Full Log-Likelihood}\label{sec:nhmm_loglik}

The log-likelihood of the entire sample of~$I$ participants is:
\begin{equation}\label{eq:full-ll}
\ell(\theta) \;=\; \sum_{i=1}^{I} \log P(\mathbf{d}_i \mid \theta)
\end{equation}
Substituting Equations \eqref{eq:marginal}, \eqref{eq:seq-ll},
\eqref{eq:forward-init}--\eqref{eq:forward-recurse}, and
\eqref{eq:transition} gives the complete expression.  The parameter vector
\[
\theta \;=\; \bigl\{\,\mu_s,\;\sigma_s,\;\pi_s,\;\boldsymbol{\beta}_j,\;c_{j,k},\;\sigma_\xi\,\bigr\}
\]
is estimated by maximizing~$\ell(\theta)$ using the Broyden--Fletcher--Goldfarb--Shanno
(BFGS) algorithm \citep{nocedal2006numerical} with 15 random initializations to
guard against local optima.  BFGS requires unconstrained parameters, so positive
quantities ($\sigma_s$, $\sigma_\xi$) are optimized in log-space and the ordered
cutpoints are reparameterized as $c_{j,0} = u_{j,0}$,
$c_{j,k} = c_{j,k-1} + \exp(u_{j,k})$ for $k \ge 1$, where the $u_{j,k}$ are
free real-valued parameters.  This guarantees the ordering constraint
$c_{j,0} < c_{j,1} < \cdots$ is satisfied at every optimization step.  The initial-state probabilities $\pi_s$ are kept on the simplex via a softmax reparameterization: the optimizer works with $N - 1$ free logits $\ell_1, \dots, \ell_{N-1}$, with $\ell_0 \equiv 0$ as the reference, and $\pi_s = \exp(\ell_s) / \sum_{k} \exp(\ell_k)$.  The emission means $\mu_s$ are unconstrained during optimization; the labeling $\mu_0 < \mu_1$ (state~0 = adherent, state~1 = non-adherent) is applied after estimation by sorting states in ascending order of their estimated emission means.
Standard errors are obtained from the numerical Hessian of the log-likelihood
evaluated at the maximum.

%% file: appendix_direct_adherence.tex
\section{Direct Dynamic Adherence Model}\label{sec:direct_adherence}

To verify that the profit-feedback patterns identified by the NHMM are not artifacts of the hidden Markov specification, we estimate a direct dynamic adherence model that tests the same hypothesis without latent-state machinery. We define a binary adherence indicator $A_{it} = \mathbf{1}[|p_{it} - p_{it}^{\mathit{AI}}| < \bar{d}]$, where $\bar{d}$ is an adherence threshold, and estimate the following linear probability model separately for each gender composition:
\begin{equation}\label{eq:direct_adherence}
    A_{it} = \alpha + \gamma_1 \, A_{i,t-1} + \gamma_2 \, \pi_{i,t-1} + \gamma_3 \, (A_{i,t-1} \times \pi_{i,t-1}) + \delta_t + \varepsilon_{it},
\end{equation}
where $A_{i,t-1}$ is lagged adherence, $\pi_{i,t-1}$ is lagged profit, and $\delta_t$ are round fixed effects. Standard errors are clustered at the cluster level. The key coefficient is $\gamma_3$: the interaction of lagged adherence and lagged profit. A positive $\gamma_3$ means that profit while adherent predicts stronger future adherence (matching the NHMM's positive $\hat{\beta}_0$ for female-only clusters), while a negative $\gamma_3$ means that profit while adherent predicts weaker future adherence (matching the negative $\hat{\beta}_0$ for male-only clusters).

Table~\ref{tab:direct_adherence} reports the results using the composition-specific median absolute deviation as the adherence threshold. The interaction coefficient $\hat{\gamma}_3$ confirms the NHMM's core finding in sign and significance for female-only clusters, where it is significantly positive ($p < 0.001$): profit earned while following the AI's recommendation predicts continued adherence. For male-only clusters, $\hat{\gamma}_3$ is significantly negative ($p = 0.009$), indicating that profit earned while following the AI predicts weaker subsequent adherence. Mixed-gender clusters show no significant interaction. The sign pattern aligns with the NHMM transition coefficients---positive for female-only, negative for male-only---and the results obtain without imposing latent states, emission distributions, or parametric transition functions.

\begin{table}[!htbp]
\centering
\caption{Direct Adherence Model: LPM Estimates}
\label{tab:direct_adherence}
\begin{threeparttable}
\begin{tabular}{lccc}
\toprule
& \multicolumn{3}{c}{\textbf{Pr(Adherence)}} \\
\cmidrule(lr){2-4}
& \textbf{Female} & \textbf{Male} & \textbf{Mixed} \\
\midrule
$A_{i,t-1}$ & $0.100^{**}$ & $0.807^{***}$ & $0.434^{***}$ \\
& $(0.044)$ & $(0.153)$ & $(0.055)$ \\[2pt]
$\pi_{i,t-1}$ & $-0.117$ & $1.287^{***}$ & $-0.002$ \\
& $(0.138)$ & $(0.408)$ & $(0.120)$ \\[2pt]
$A_{i,t-1} \times \pi_{i,t-1}$ & $1.802^{***}$ & $-2.208^{***}$ & $0.233$ \\
& $(0.244)$ & $(0.849)$ & $(0.338)$ \\[2pt]
\midrule
Round FE & Yes & Yes & Yes \\
Observations & 1{,}085 & 1{,}198 & 1{,}404 \\
Clusters & 15 & 15 & 19 \\
Adherence threshold & 0.05 & 0.03 & 0.14 \\
\bottomrule
\end{tabular}
\begin{tablenotes}
\footnotesize
\item Notes. Linear probability models estimated separately by gender composition on the treatment-group subsample. The dependent variable is a binary adherence indicator $A_{it} = \mathbf{1}[|p_{it} - p_{it}^{\mathit{AI}}| < \bar{d}]$. The adherence threshold is the composition-specific median absolute deviation from the AI recommendation. Standard errors clustered at the cluster level in parentheses. $^{*}\,p<0.10$; $^{**}\,p<0.05$; $^{***}\,p<0.01$.
\end{tablenotes}
\end{threeparttable}
\end{table}

%% file: rationale_lpm_appendix.tex
\section{Systematic Rationale Coding and Rationale-Based LPMs}\label{sec:rationale_lpm_appendix}

This appendix describes the text-based analysis used in Section~\ref{sec:mechanism} to validate the behavioral interpretation of the NHMM transition estimates. After each pricing decision, participants entered a free-text rationale explaining the basis for their price. We classify each treatment-group seller-round rationale into one of five mutually exclusive categories: \textit{trust-AI}, \textit{own-strategy}, \textit{market-factors}, \textit{anchoring}, and \textit{other}. The coding is deductive and theory-driven. A rationale is coded as \textit{trust-AI} when it explicitly references following, trusting, relying on, or agreeing with the AI recommendation. A rationale is coded as \textit{own-strategy} when it references the seller's own judgment, skill, model, analysis, or independent pricing strategy. \textit{Market-factors} captures rationales based on competitors, demand, market share, customer response, or price trends; \textit{anchoring} captures maintaining the previous price or making small inertial adjustments; \textit{other} captures unclear or unclassifiable text.

We use an LLM-assisted coding procedure with a fixed codebook and structured output, implemented with GPT-5.4 mini. The model observes only the rationale text and the category definitions; it does not observe the participant's gender, market composition, price, profit, or subsequent behavior. The output for each rationale is a category label, a confidence score, and a one-sentence coding justification. The coding file is then merged back to the same treatment-group seller-round panel used in the direct dynamic adherence model (Appendix~\ref{sec:direct_adherence}).

\begin{lstlisting}[style=promptstyle,
                   float=htbp,
                   caption={LLM prompt for systematic rationale coding.},
                   label={lst:rationale_coding_prompt}]
(*@\textbf{System instruction}@*)
You are a research assistant coding participant rationales from a laboratory pricing experiment. In this experiment, sellers in an online marketplace set prices each round and optionally received an AI-generated price recommendation. After each price submission, participants wrote a free-text rationale explaining their pricing decision.

Your task: classify each rationale into exactly ONE of the following five categories.

## Coding Categories

### trust_ai
The rationale explicitly references trusting, following, relying on, or agreeing with the AI's recommendation. This includes:
- Direct mentions of AI, algorithm, or the recommendation ("follow AI", "AI suggestion", "trusted the recommendation")
- Statements that the AI's price is a good guide ("AI seems right", "going with what AI says")
- In Chinese: references to AI(*@\zh{建议}@*), AI(*@\zh{推荐}@*), (*@\zh{跟随}@*)AI, (*@\zh{按照}@*)AI, etc.
Do NOT code as trust_ai if the participant merely set a price close to AI without mentioning it.

### own_strategy
The rationale references the participant's own judgment, skill, analysis, personal strategy, or independent decision-making. This includes:
- Claims of personal insight ("my analysis", "I think", "my strategy", "based on my model")
- Expressions of confidence in own pricing ability ("I know the market", "my approach works")
- Deliberate deviation framed as own choice ("trying my own price", "experimenting")
- In Chinese: (*@\zh{我的策略}@*), (*@\zh{我认为}@*), (*@\zh{我觉得}@*), (*@\zh{自己的判断}@*), etc.

### market_factors
The rationale references external market conditions, competitor behavior, supply/demand, or price trends. This includes:
- Competitor price references ("others are pricing lower", "undercut competitors", "match market")
- Demand/market share reasoning ("capture more customers", "market share dropping")
- Price trend observations ("prices are rising", "market stabilizing")
- In Chinese: (*@\zh{市场}@*), (*@\zh{竞争}@*), (*@\zh{价格趋势}@*), (*@\zh{客户}@*), (*@\zh{市场份额}@*), etc.

### anchoring
The rationale references maintaining the previous price, making minimal adjustments, or inertia. This includes:
- Explicit price maintenance ("keep same price", "no change", "hold steady")
- Small adjustments without clear reasoning ("slightly higher", "adjust a bit")
- References to prior round price as anchor ("same as last round")
- In Chinese: (*@\zh{保持}@*), (*@\zh{不变}@*), (*@\zh{维持}@*), (*@\zh{微调}@*), etc.

### other
Use this category ONLY when the rationale is:
- Too short or vague to classify (single words like "ok", "test", "try", numbers only)
- Genuinely ambiguous between multiple categories with no dominant theme
- Irrelevant or nonsensical content

## Important Rules
1. Code the PRIMARY rationale. If multiple themes are present, choose the dominant one.
2. Code what the participant WROTE, not what you infer from the price data.
3. Apply the same standards regardless of the participant's gender (which you do not know).
4. Handle Chinese and English rationales with equal rigor.
5. Very short rationales (1-2 words) should be coded if the meaning is clear (e.g., "ai" (*@$\rightarrow$@*) trust_ai, "undercut" (*@$\rightarrow$@*) market_factors), otherwise (*@$\rightarrow$@*) other.

(*@\textbf{User message template}@*)
Classify this pricing rationale:

"{rationale}"
\end{lstlisting}

To test whether the textual rationales support the proposed mechanisms, we estimate two linear probability models. The first dependent variable is
\[
    \textit{TrustAI}_{it} = \mathbf{1}\{\text{rationale}_{it} \text{ is coded as trust-AI}\},
\]
and the second is
\[
    \textit{OwnStrategy}_{it} = \mathbf{1}\{\text{rationale}_{it} \text{ is coded as own-strategy}\}.
\]
For each dependent variable, we estimate separately by gender composition:
\begin{equation}\label{eq:rationale_lpm}
    Y_{it} = \alpha
    + \gamma_1 A_{i,t-1}
    + \gamma_2 \pi_{i,t-1}
    + \gamma_3 (A_{i,t-1} \times \pi_{i,t-1})
    + \delta_t
    + \varepsilon_{it},
\end{equation}
where $A_{i,t-1}$ is lagged adherence to the AI recommendation, $\pi_{i,t-1}$ is lagged profit, and $\delta_t$ are round fixed effects. We define lagged adherence using the composition-specific median absolute deviation from the AI recommendation: $0.05$ in female-only markets, $0.03$ in male-only markets, and $0.14$ in mixed-gender markets. Standard errors are clustered at the cluster level.

The coefficient of interest is $\gamma_3$. In the trust-AI model, a positive $\gamma_3$ means that profit earned while following the AI makes a seller more likely to cite the AI as the basis for the next pricing decision. This is the textual implication of learned trust. In the own-strategy model, a positive $\gamma_3$ means that profit earned while following the AI makes a seller more likely to cite their own expertise or strategy in the next rationale. This is the textual implication of self-serving attribution.

\begin{table}[!htbp]
\centering
\caption{Rationale-Based Linear Probability Models}
\label{tab:rationale_lpm}
\begin{threeparttable}
\begin{tabular}{lcccccc}
\toprule
& \multicolumn{3}{c}{\textbf{Pr(Trust-AI Rationale)}} & \multicolumn{3}{c}{\textbf{Pr(Own-Strategy Rationale)}} \\
\cmidrule(lr){2-4} \cmidrule(lr){5-7}
& \textbf{Female} & \textbf{Male} & \textbf{Mixed} & \textbf{Female} & \textbf{Male} & \textbf{Mixed} \\
\midrule
$A_{i,t-1}$ & $-0.028$ & $0.025$ & $0.109^{**}$ & $-0.069$ & $-0.201^{*}$ & $-0.017$ \\
& $(0.102)$ & $(0.067)$ & $(0.045)$ & $(0.081)$ & $(0.109)$ & $(0.053)$ \\[2pt]
$\pi_{i,t-1}$ & $-0.130$ & $-0.282^{***}$ & $-0.182$ & $0.001$ & $0.233$ & $0.110$ \\
& $(0.147)$ & $(0.094)$ & $(0.122)$ & $(0.302)$ & $(0.149)$ & $(0.160)$ \\[2pt]
$A_{i,t-1} \times \pi_{i,t-1}$ & $1.557^{**}$ & $-0.085$ & $-0.077$ & $-0.069$ & $3.342^{***}$ & $-0.385$ \\
& $(0.759)$ & $(0.226)$ & $(0.266)$ & $(0.418)$ & $(0.550)$ & $(0.284)$ \\[2pt]
\midrule
Round FE & Yes & Yes & Yes & Yes & Yes & Yes \\
Observations & 1{,}085 & 1{,}198 & 1{,}404 & 1{,}085 & 1{,}198 & 1{,}404 \\
Clusters & 15 & 15 & 19 & 15 & 15 & 19 \\
Adherence threshold & 0.05 & 0.03 & 0.14 & 0.05 & 0.03 & 0.14 \\
\bottomrule
\end{tabular}
\begin{tablenotes}
\footnotesize
\item Notes. Linear probability models estimated separately by gender composition on the same treatment-group seller-round panel used in Appendix~\ref{sec:direct_adherence}. The dependent variable in Columns 1--3 equals one if the rationale is coded as \textit{trust-AI}; the dependent variable in Columns 4--6 equals one if the rationale is coded as \textit{own-strategy}. $A_{i,t-1}$ is a binary lagged-adherence indicator based on the composition-specific median absolute deviation from the AI recommendation. Standard errors clustered at the cluster level in parentheses. $^{*}\,p<0.10$; $^{**}\,p<0.05$; $^{***}\,p<0.01$.
\end{tablenotes}
\end{threeparttable}
\end{table}

The results align with the mechanism proposed in the main text. In female-only markets, the interaction between lagged adherence and lagged profit significantly predicts trust-AI rationales ($\hat{\gamma}_3 = 1.557$, $p = 0.040$), but the same interaction does not predict own-strategy rationales. In male-only markets, the corresponding interaction significantly predicts own-strategy rationales ($\hat{\gamma}_3 = 3.342$, $p < 0.001$), but does not predict trust-AI rationales. Mixed-gender markets show neither pattern. Thus the text analysis corroborates the interpretation that profitable adherence produces learned trust in female-only markets and self-serving attribution in male-only markets.

%% file: appendix_decomposition.tex
\section{Decomposition of AI-Signal and Human-Response Heterogeneity}\label{sec:decomposition}

The NHMM in Section~\ref{sec:mechanism} is estimated on the treatment group, where the AI adapts its recommendations to each market's pricing history. A composition effect on adherence could therefore reflect two channels: (i) human sellers in different compositions respond differently to similar recommendations (human-response heterogeneity), or (ii) the adaptive AI generates recommendation sequences that differ systematically across compositions, and sellers respond similarly to different signals (AI-signal heterogeneity). To isolate the first channel, we control nonparametrically for the recommendation signal.

We estimate
\begin{equation}\label{eq:decomposition}
|d_{it} - r_{it}| = \alpha + \boldsymbol{\beta}\,\mathbf{Comp}_i + \boldsymbol{\gamma}\,\mathbf{1}[\text{RecDecile}_{it}] + \delta\,\text{RecGap}_{it} + \boldsymbol{\theta}\,\mathbf{X}_{i,t-1} + \varepsilon_{it},
\end{equation}
where $d_{it}$ is the price set by seller $i$ in round $t$, $r_{it}$ is the AI recommendation, $\mathbf{Comp}_i$ is a vector of composition indicators (male-only is the reference category), $\mathbf{1}[\text{RecDecile}_{it}]$ is a set of recommendation-decile fixed effects that absorb any level-specific effect of the recommendation, $\text{RecGap}_{it} = r_{it} - \bar{p}^{\text{ctrl}}_{\text{comp}(i)}$ is the gap between the recommendation and the composition-specific control-group baseline price, and $\mathbf{X}_{i,t-1}$ includes lagged profit, lagged own price, lagged market share, lagged market average price, and round number. Standard errors are clustered at the cluster level.

The recommendation-decile fixed effects are the key control: they partition the recommendation distribution into ten equally sized bins, so the composition coefficients $\boldsymbol{\beta}$ capture only within-bin differences in how sellers respond to recommendations of similar magnitude. If the composition effect were driven entirely by the AI sending different signals to different compositions, $\boldsymbol{\beta}$ would be zero once the signal is absorbed.

Table~\ref{tab:decomposition_m4} reports the results. The sample comprises 3,687 seller-round observations from 147 treated sellers in 49 clusters, after dropping the first round (which lacks lagged covariates).

\begin{table}[!htbp]
\centering
\caption{Decomposition Test: Absolute Deviation from AI Recommendation (M4)}
\label{tab:decomposition_m4}
\begin{threeparttable}
\begin{tabular}{lcccc}
\toprule
& Coef. & SE & $z$ & $p$ \\
\midrule
\multicolumn{5}{l}{\textit{Composition (ref: male-only)}} \\[3pt]
\quad Female-only     & $-0.140$ & $0.041$ & $-3.39$ & $<0.001$ \\
\quad Mixed-gender    & $-0.065$ & $0.049$ & $-1.33$ & $0.185$ \\[6pt]
\multicolumn{5}{l}{\textit{Market history controls}} \\[3pt]
\quad Lag profit         & $-0.577$ & $0.130$ & $-4.42$ & $<0.001$ \\
\quad Lag own price      & $0.200$  & $0.079$ & $2.54$  & $0.011$ \\
\quad Lag market share   & $0.540$  & $0.109$ & $4.95$  & $<0.001$ \\
\quad Lag market avg.\ price & $-0.206$ & $0.069$ & $-2.98$ & $0.003$ \\
\quad Round number       & $-0.004$ & $0.001$ & $-3.63$ & $<0.001$ \\
\quad Prior GPT Trust    & $0.017$  & $0.018$ & $0.93$  & $0.350$ \\[6pt]
\quad Rec.--baseline gap & $0.361$  & $0.117$ & $3.10$  & $0.002$ \\[3pt]
\midrule
Rec.\ decile FE & \multicolumn{4}{c}{Yes (9 indicators)} \\
$N$ & \multicolumn{4}{c}{3,687} \\
Clusters & \multicolumn{4}{c}{49} \\
$R^2$ & \multicolumn{4}{c}{0.166} \\
\bottomrule
\end{tabular}
\begin{tablenotes}
\footnotesize
\item Notes. OLS estimates of Equation~\eqref{eq:decomposition}. The dependent variable is the absolute deviation $|d_{it} - r_{it}|$. Recommendation-decile fixed effects partition the AI recommendation into ten equally sized bins; their coefficients are omitted for brevity. Standard errors (SE) are robust to clustering at the cluster level. The sample includes all treated sellers in rounds 2--30 (round 1 is dropped because lagged covariates are undefined).
\end{tablenotes}
\end{threeparttable}
\end{table}

Female-only clusters deviate 0.140 less from the AI recommendation than male-only clusters within the same recommendation decile ($p < 0.001$). This effect survives the inclusion of recommendation-decile fixed effects, which absorb any composition-specific pattern in what the AI recommends. Mixed-gender clusters do not differ significantly from male-only clusters ($p = 0.185$). These results establish that the adherence divergence documented in the main text reflects heterogeneous human responses to similar AI signals, not merely heterogeneous AI signals across compositions.